\documentclass[11pt]{article}

\usepackage[a4paper,margin=1in]{geometry}
\usepackage{graphicx}
\usepackage{authblk}
\usepackage[numbers,sort&compress]{natbib}
\usepackage{hyperref}
\usepackage{lineno}
\usepackage{multirow}
\usepackage{amsmath,amssymb,amsfonts}
\usepackage{amsthm}
\usepackage{mathrsfs}
\usepackage[title]{appendix}
\usepackage{xcolor}
\usepackage{textcomp}
\usepackage{manyfoot}
\usepackage{booktabs}
\usepackage{algorithm}
\usepackage{algorithmicx}
\usepackage{algpseudocode}
\usepackage{listings}
\usepackage{tabularx}
\usepackage{orcidlink} 
\usepackage{soul}
\providecommand{\bmhead}[1]{\subsection*{#1}}

\title{Joint sparse coding and temporal dynamics support context reconfiguration}

\author[1,2,3]{Qianqian Shi}

\author[4]{Yue Che}

\author[1]{Faqiang Liu}

\author[1,2,3]{Hongyi Li}

\author[4]{Mingkun Xu}

\author[5]{Sandra Reinert}

\author[6]{Pieter M. Goltstein}

\author[1,2,3]{Rong Zhao}

\author[1,2,3,*]{Luping Shi}

\affil[1]{Center for Brain-Inspired Computing Research (CBICR), Department of Precision Instrument, Tsinghua University, Beijing, China}

\affil[2]{Optical Memory National Engineering Research Center, Tsinghua University, Beijing, China}

\affil[3]{IDG/McGovern Institute for Brain Research, Tsinghua University, Beijing, China}

\affil[4]{Guangdong Institute of Intelligence Science and Technology, Hengqin, China}

\affil[5]{Sainsbury Wellcome Centre, University College London, London, UK}

\affil[6]{Max Planck Institute for Biological Intelligence, Martinsried, Germany}

\date{}

\begin{document}

\maketitle

\begin{abstract}
Adaptive behavior requires the brain to transition between distinct contexts while maintaining representations of prior experience. The ability to reconfigure neural representations without erasing previously acquired knowledge is central to learning in dynamic environments, yet the neural mechanisms that support this balance remain unclear. Understanding these mechanisms is also critical for addressing catastrophic forgetting in artificial systems designed for lifelong learning. Here, we identify joint sparse coding and temporal dynamics in both the mouse medial prefrontal cortex (mPFC) and computational networks as mechanisms that help preserve prior representations during context transitions. Specifically, sparsity in context-dependent representations reduces cross-context interference, whereas temporal dynamics within the network activity further enhance context separability across time. Strikingly, networks endowed with both properties, such as spiking neural networks, exhibit improved retention during lifelong learning without auxiliary heuristics. These findings establish joint sparse coding and temporal dynamics as a core mechanism supporting flexible context reconfiguration in lifelong learning and, through their activity-constraining nature, as an energy-efficient architectural principle for stable adaptation. Together, they provide a mechanistic framework for understanding how the brain preserves prior knowledge while flexibly adapting to new contexts.
\end{abstract}

\section*{Introduction}

A fundamental question in neuroscience is how the brain supports flexible yet stable learning across changing contexts, rapidly reconfiguring behavior while preserving previously acquired knowledge. Maintaining this stability–flexibility balance is essential for learning and adaptive behavior. Converging evidence implicates a distributed prefrontal control network in mediating rule and context transitions \cite{Sakai2008, Coolsetal2004}. In rule-switching paradigms, population activity in medial prefrontal cortex encodes the currently relevant task set, whereas functionally specialized subregions, including anterior cingulate areas, contribute to conflict monitoring and performance evaluation that support updating of established rules \cite{MillerCohen2001, Botvinicketal2004}. More broadly, contextual processing engages coordinated hippocampal–prefrontal interactions, consistent with a circuit motif in which prefrontal control enables rapid reconfiguration without wholesale erasure of prior policies \cite{PrestonEichenbaum2013, Placeetal2016, SigurdssonDuvarci2016}.

However, although these circuit-level findings provide a strong foundation for understanding rule and context transitions, they do not yet identify the population-level mechanism by which neural populations learn across sequentially encountered tasks or contexts. A central unresolved question is how neural ensembles remain sufficiently stable to preserve prior representations, yet sufficiently flexible to be selectively reconfigured when contexts change, thereby limiting interference during sequential learning.

A related challenge arises in artificial intelligence. Despite major advances in perception and decision-making \cite{LeCun2015}, standard artificial neural networks often fail when tasks or contexts are encountered sequentially, a limitation commonly referred to as catastrophic forgetting \cite{parisi2019continual, de2021continual}, frequently accompanied by substantial memory and energy demands \cite{shi2025hybrid}. At the same time, neuroscience and artificial intelligence have long advanced through reciprocal exchange: biological principles have inspired core developments in machine learning, including convolutional architectures \cite{HubelWiesel1962, LeCun1998}, hierarchical deep representations \cite{Fukushima1980, Yamins2014}, attention mechanisms \cite{DesimoneDuncan1995, Vaswani2017} and replay-based lifelong learning \cite{van2020brain}, whereas computational frameworks have sharpened hypotheses about neural computation and provided quantitative tools for interrogating brain circuits \cite{Richards2019, Kriegeskorte2015}. Together, these links raise the possibility that population-level coding principles in brain-like networks may inform solutions to catastrophic interference in artificial systems. In turn, computational models provide a tractable framework for testing and refining hypotheses about the neural mechanisms that support lifelong learning, also known as continual learning.

Two biological properties are especially relevant to this problem: sparse coding and temporal dynamics. Sparse coding, in which only a subset of neurons in a population is strongly recruited by a given input or context, has long been proposed as an efficient coding strategy in neuroscience \cite{Barlow1961,Field1987}. It has been studied most extensively in sensory systems, where sparse representation has been proposed as a solution to the trade-off between resolving individual sensory inputs and generalizing across relevant features \cite{RollsTreves1990}. Beyond sensory systems, sparse coding principles have also been described in frontal cortex \cite{Abeles1990}, a view supported by more recent studies demonstrating that only small fractions of neurons are selective for any given feature or task context \cite{Roy2010,Rikhye2018}. These observations suggest that sparse population recruitment may be well suited to resolving the trade-off between faithfully encoding specific contexts and enabling generalization to novel contexts, especially within the lifelong learning regimes that characterize natural behavior.

Temporal dynamics constitute a second fundamental property of biological neural circuits. They emerge from membrane and synaptic time constants together with recurrent interactions, such that neural responses depend on recent history rather than solely on instantaneous input \cite{Buonomano2009StateDependent}. As a result, neural representations are shaped not only by which neurons are recruited, but also by how their activity unfolds across successive time steps, extending coding beyond a purely instantaneous population pattern. At the population level, such temporal structure can coexist with stable, readout-relevant representations despite dynamic responses in individual neurons \cite{Murray2017StablePopulation}. Recent work further suggests that temporally structured cortical activity can enhance representational discriminability and support more stable sensory representations over time \cite{Zhu2025TemporalCoding}. By distributing information across successive time steps, temporal dynamics may therefore not only stabilize context-related representations, but also provide an additional temporal dimension through which neural populations can organize and differentiate their activity.

Although sparse recruitment and temporal dynamics have each been studied extensively in other settings, their roles in lifelong learning—and in particular their potential interaction in supporting flexible context reconfiguration—remain unresolved. We therefore asked whether these two properties together define a population-level mechanism by which neural populations can incorporate new contexts while limiting interference with previously acquired representations.

Here, focusing on mouse mPFC, a key locus for cognitive flexibility and rule updating \cite{Sakai2008}, we find that context transitions are associated with a sparse and temporally structured population code. At the population level, different contexts recruit distinct yet partially overlapping neuronal ensembles, such that neurons strongly active in one context are often weakly active or silent in the other, consistent with sparse context coding. At the temporal level, context decoding depends on both the length of the temporal integration window and the temporal ordering of neural activity patterns, indicating that temporal dynamics enhance the discriminability of contextual representations.

Extending these observations to computational models, we find that spiking neural networks (SNNs), whose dynamics naturally impose sparse recruitment and temporal structure \cite{Gerstner2002, Maass1997}, show enhanced resistance to catastrophic forgetting relative to conventional artificial neural networks (ANNs) under lifelong-learning protocols. To dissect the underlying principle, we implemented sparse coding and state-dependent temporal dynamics in ANNs independently and in combination. This reveals cooperative roles: sparse coding reduces cross-context interference by partitioning activity into partially context-selective ensembles, whereas temporal dynamics are not effective in isolation but, when coupled with sparse recruitment, further separate context-dependent activity across time. Together, these mechanisms increase context discriminability, reduce interference, and recapitulate the improved performance and robustness observed in SNNs.

Furthermore, beyond accuracy, sparse, event-driven dynamics can also confer efficiency advantages relevant to biological constraints and neuromorphic implementations \cite{orchard2021efficient, Indiveri2011}. Under a biologically inspired local plasticity learning strategy \cite{journe2022hebbian} rather than backpropagation (BP), SNNs retain performance advantages while substantially reducing memory footprint and computational demands. These properties align with the operating principles of neuromorphic
hardware, where asynchronous event-driven computation supports low-power, scalable
lifelong learning \cite{pei2019towards, gonzalez2024spinnaker2}.

Together, our findings identify sparse coding and temporal dynamics as cooperative mechanisms that enable stable integration of new experience while preserving established knowledge, supporting adaptive context reconfiguration. Their inherent metabolic efficiency further underscores its suitability as a general principle for sustained learning in both biological and artificial systems.

\section*{Results}

\textbf{Sparse coding and temporal dynamics in medial prefrontal cortex support context reconfiguration}\\
Flexible behavior requires the ability to adapt to new behavioral policies or task rules without disrupting previously acquired knowledge. Here, we propose that context reconfiguration is supported by sparse coding and intrinsic temporal dynamics, whose interaction in particular promotes the emergence of more independent subnetworks across contexts (Fig. 1).

At the population level, context representations can range from local to dense coding regimes. Sparse coding represents an intermediate organization in which different contexts recruit partially non-overlapping neuronal ensembles, while a subset of neurons is shared across contexts (as shown in Fig.1b). Such ensemble structure allows context-dependent representations to remain distinct while preserving shared units across contexts \cite{Spanne_Jorntell_2015}. 

At the temporal level, intrinsic neuronal dynamics further shape context representations by extending neural responses across successive time points. As illustrated in Fig.1c, synaptic integration and recurrent interactions generate history-dependent activity patterns that are not confined to instantaneous activation \cite{Buonomano2009StateDependent,Wang2001,murray2014hierarchy}. Thus, context coding depends not only on which neurons are recruited, but also on how their activity unfolds over time. Accordingly, neurons with similar mean firing rates across contexts may nevertheless differ in their temporal response structure, such that context information becomes more discriminable when responses are evaluated over extended temporal windows rather than at single time points \cite{Zhu2025TemporalCoding}. Together, these considerations suggest a working framework in which different contexts recruit partially overlapping but non-identical neuronal ensembles, and in which temporal dynamics further enhance the discriminability of those context representations over extended time windows (Fig.1d). Such spatiotemporal organization may help reduce interference between contexts by biasing population activity toward more distinct functional subnetworks.

To test these predictions, we analyzed population recordings from the mouse medial prefrontal cortex during a rule-based Go/NoGo categorization task \cite{reinert2021mouse}. Head-fixed mice were trained to categorize sinusoidal grating stimuli according to a single task rule, in which either spatial frequency or orientation determined category membership (Fig.2a, 2b, 2c). Training proceeded in a staged manner, beginning with discrimination between a small number of exemplars and gradually expanding to a larger stimulus set (up to 36 gratings) until stable performance was achieved. Following rule acquisition, the task context was switched by making the previously irrelevant stimulus feature relevant and vice versa. After each switch, mice were retrained stepwise under the new rule, starting with a small subset of stimuli before progressively returning to the full stimulus set. Behavioral performance typically dropped immediately after the switch and recovered over several sessions, indicating the need for context-dependent relearning.

We defined the spatial-frequency rule as context X and the orientation rule as context Y. To control for motor-response effects, we analyzed only response-matched trials, including stimuli that were Go in both contexts and stimuli that were NoGo in both contexts. We then compared trial-averaged inferred firing rates between contexts. Many neurons ranked among the top 100 in one context showed much lower firing rates in the other, whereas only a smaller subset ranked among the top 100 in both contexts (Fig.2d, 2e).

To quantify ensemble overlap across contexts, neurons were ranked by inferred firing rate averaged across trials within each context. We then calculated the overlap between the top 100 neurons in the two contexts. The resulting cross-context overlap fraction was 0.32 (Fig. 2f), which was substantially below the shuffle-derived chance level (0.61 $\pm$ 0.04). This result suggests that context-specific activity recruits neuronal ensembles that are only partially shared across contexts.

We next computed a context tuning index (CxTI), adapted from the category tuning index \cite{reinert2021mouse}, to quantify differential neuronal activity between contexts (Extended Data Fig.1a). We then used CxTI-selected neurons to train a linear support vector machine (SVM) to decode context identity. The classifier reliably discriminated context X from context Y, achieving a mean decoding accuracy of approximately 82.58\%, whereas decoding with shuffled context labels remained at chance level (approximately 50.1\%; Fig.2g; ***p $<$ 0.001).Decoding based on CxTI-selected neurons also significantly outperformed decoding based on randomly selected neurons (Extended Data Fig.1b; ***p $<$ 0.001), indicating that CxTI effectively captures context-informative neural populations.

We next examined how temporal structure in stimulus-evoked mPFC population activity contributes to context encoding. We systematically varied the temporal integration window (T, in frames) using two sampling strategies: contiguous temporal segments and randomly selected discrete time points. Context identity was decoded from population activity using a linear SVM. Given an imaging frame of 10 Hz, T ranged from 2 to 14 frames.

In both sampling strategies, decoding accuracy increased monotonically with longer integration windows (Fig.2h, Extended Data Fig.1c),  indicating that incorporating more temporal information improves the separability of context representations and thereby facilitates context decoding. 

Notably, when controlling for the number of sampled time points (that is, matched T), decoding accuracy was significantly higher for contiguous temporal segments than for discrete time points (Fig.2i; *p $=$ 0.0426). In both conditions, decoding used the same neurons and trials, with time points subsampled from within the stimulus window (2 to N$-$1 frames) rather than the full temporal sequence. To further isolate the contribution of temporal continuity, we performed an additional control analysis using the same sampled time points but shuffling their temporal order. Under this matched comparison, decoding accuracy was again significantly higher for the original contiguous sequence than for the temporally shuffled sequence (Extended Data Fig.1d; *p $=$ 0.0114), indicating that the gain in decoding cannot be explained solely by the number of sampled frames or the total amount of activity information, but depends on the preserved temporal structure of the response.

Together, these results indicate that, beyond the total amount of temporal information, the preservation of local temporal continuity, which captures the dynamics linking consecutive time points, increases the discriminability of context representations and improves context decoding accuracy.
\par\vspace{0.5\baselineskip}
\noindent\textbf{Built-in threshold-induced sparsity and temporal dynamics enable superior lifelong learning in spiking neural networks.}\\
To directly test whether sparse and temporally structured coding is sufficient to support context-dependent computation, we evaluated models under a lifelong-learning setting. In this standardized framework, a shared network must sequentially adapt to different task rules following context transitions (Fig.3a). 

We compared spiking neural networks with conventional artificial neural networks. SNNs naturally combine sparse, event-driven population activity with intrinsic temporal dynamics arising from membrane-potential integration, decay, and reset \cite{Gerstner2002, Maass1997}. In contrast, ANNs produce dense, continuous activations without intrinsic state dynamics (Fig.3b, 3c; Table 1). For fair comparison, both architectures were capacity-matched and trained using the same backpropagation-based procedure, isolating the contribution of representational coding rather than optimization differences.

A single-hidden-layer architecture was served as the default configuration in most experiments, as increasing network depth from one to two hidden layers consistently reduced performance in both SNN and ANNs under sequential training (Extended Data Fig.2c), in line with prior observations \cite{dohare2024loss}. For completeness, selected analyses were also conducted using a two-hidden-layer architecture (e.g., Fig.3e-g), as indicated in the figure panels.

Within the SNN framework, we employed spiking neurons based on the ternary leaky integrate-and-fire (TLIF) model \cite{guo2024ternary}, in which neurons emit ternary-valued outputs ($+$1, 0, $-$1), corresponding to excitatory, silent, and inhibitory states. To determine how the spatial extent of integration-and-fire dynamics influences lifelong learning, we implemented two configurations. In the single integration-and-fire (1 IF) configuration, spiking dynamics were applied only prior to the hidden layer. In the dual integration-and-fire (2 IF) configuration, spiking dynamics were applied both prior to the hidden layer and prior to the classifier, thereby extending threshold gating and temporal state evolution throughout the network.

Applying integration-and-fire dynamics across both layers (2 IF) consistently improved lifelong-learning performance relative to the 1 IF configuration (Extended Data Fig.2d), suggesting that network-wide deployment of sparse, temporally dynamic units more effectively enhances incremental learning performance. Accordingly, unless otherwise noted, subsequent SNN experiments employ the 2 IF configuration.

We evaluated these architectures across task-incremental (TIL), domain-incremental (DIL), and class-incremental learning (CIL) settings \cite{van2022three}. Consistent with the hypothesis that sparse, temporally structured coding promotes representational stability, SNNs outperformed capacity-matched ANNs across all evaluated lifelong-learning scenarios, and this advantage remained evident even when ANNs were scaled to higher capacity (Fig.3d; Extended Data Fig.2a, 2b). Because class-incremental learning is widely regarded as the most stringent of these settings \cite{van2022three}, we focus our main analyses on this regime. Under the class-incremental setting, the performance gap between architectures increased with the number of incremental stages (Fig.3e). By the end of training, ANNs exhibited a pronounced decline in accuracy on earlier stages, consistent with catastrophic forgetting, whereas SNNs retained substantially more previously acquired knowledge while incorporating new classes (Fig.3f, 3g). 

This improved retention was accompanied by lower overlap of context-recruited neurons in SNNs than in ANNs (Fig.3h, 3i), indicating that SNNs form more segregated context-specific subnetworks, whereas ANNs rely on more overlapping neuronal populations that are more susceptible to representational interference.

An important open question is whether enhanced context-specific representational separation compromises transfer across tasks. Previous work has suggested a potential trade-off between representational separation and transfer, whereby stronger task segregation may reduce generalization across related tasks \cite{holton2025humans}. Given that SNNs exhibited reduced inter-task overlap, we explicitly tested whether this increased separation impairs transfer.

To address this question, we evaluated both models on the permuted MNIST (pMNIST) benchmark \cite{lecun2002gradient}, which consists of four task groups characterized by high within-group similarity and larger differences across groups (Extended Data Fig.3a, 3b). During sequential learning, ANNs often achieved higher accuracy on newly introduced stages when trained for the same number of epochs (Extended Data Fig.3c), consistent with faster performance gains per epoch. In contrast, SNNs exhibited superior retention of previously learned tasks across stages within the same task group, reflected by significantly reduced forgetting rates (Fig.3j, ***p $<$ 0.001). 

To directly quantify transfer, we defined transfer efficiency as the ratio between performance on transferred tasks and that obtained through direct training. Within each task group, transfer efficiency was comparable between SNNs and ANNs, with no significant difference observed (Fig.3k, n.s.). Thus, although SNNs exhibited stronger representational separation and improved retention, this did not come at the expense of reduced within-group transfer.
\par\vspace{0.5\baselineskip}
\noindent\textbf{Distributed sparsity enhances performance in lifelong learning}\\
In the class-incremental setting, varying the spiking threshold systematically modulated SNN performance (Extended Data Fig.5a), indicating a tight coupling between sparsity level and lifelong-learning outcomes. Because spiking neural networks simultaneously introduce sparsity and temporal dynamics, we next sought to isolate the contribution of threshold-gated sparsity independent of temporal state dynamics.

To this end, we introduced threshold-induced sparsity into ANNs using a thresholded activation rule (see Methods), such that units produced real-valued outputs only when pre-activations exceeded a threshold and were otherwise silent. This manipulation imposed sparse, context-dependent recruitment while eliminating temporal integration effects (Fig.4a).

To test whether the spatial extent of sparsity determines its efficacy, we implemented two architectures differing in the scope of threshold gating. In the local-sparsity condition, thresholding was applied only at the input-to-hidden transformation (1 IF). In the network-wide sparsity condition, threshold gating was applied at both input-to-hidden and hidden-to-output transformations (2 IF), thereby constraining activity throughout the network. 

Local sparsity produced little to no improvement—and in some regimes even reduced performance—relative to baseline ANNs (Fig.4b, 4d). In contrast, network-wide sparsity led to a significant enhancement in lifelong-learning performance (Fig.4c, 4e). Together, these results indicate that sparsity improves lifelong learning only when it is sufficiently strong and distributed to partition population activity into context-dependent subnetworks, rather than when confined to a local transformation.

At the population level, however, neither local nor network-wide sparsity substantially reduced inter-context representational overlap (Fig.4f; Extended Data Fig.2f, 6). Although network-wide sparsity improved performance, overlap patterns remained largely similar across sparsity configurations. This dissociation indicates that distributed sparsity alone is insufficient to markedly enhance population-level context separation, and suggests that temporal dynamics are required to further increase the discriminability of context-specific representations.
\par\vspace{0.5\baselineskip}
\noindent\textbf{State-dependent temporal dynamics cooperate with sparsity to enhance context separation in lifelong learning}\\
Having established that spatially distributed sparsity alone is insufficient to fully reproduce the context segregation observed in SNNs, we next asked whether intrinsic temporal dynamics provide an additional mechanism for separating context representations during sequential learning..

To directly test this necessity, we collapsed network computation to a single time step, thereby eliminating membrane-potential integration and spike-triggered reset. Under this manipulation, SNNs were reduced to a binary, stateless architecture, resulting in a marked decline in lifelong-learning performance (Extended Data Fig.5b). This results demonstrates that intrinsic temporal dynamics are essential for the robustness of SNNs under sequential training.

We next asked whether temporal dynamics can improve performance in non-spiking networks. To this end, we incorporated TLIF-like state dynamics into ANNs by introducing decay and event-triggered reset (see Methods), such that each unit maintained a history-dependent internal state over time (Fig.5a). In the absence of population-level sparsity, however, temporal dynamics alone failed to improve lifelong-learning performance (Extended Data Fig.5c), indicating that temporal state evolution is not sufficient in isolation.

By contrast, when temporal dynamics were combined with sparsity, their impact became pronounced. Under the local sparsity configuration, increasing the temporal integration length significantly improved performance, approaching that of SNNs (Fig.5b, 5c), and produced levels of context independence comparable to SNNs (Fig.5f). Under the network-wide sparsity configuration, temporal dynamics yielded only modest changes in mean accuracy (Fig.5d) but substantially increased robustness across thresholds by reducing sensitivity to sparsity strength (Fig.5e; Fig.4e).

To further dissect the mechanism, we ablated the reset component of the temporal dynamics. Removing reset, and thereby eliminating the refractory-like constraint on activity \cite{Gerstner2014}, significantly impaired lifelong-learning performance (Fig.5c, 5d). One possible interpretation is that reset promotes temporal sparsification by suppressing the repeated recruitment of recently active units. Its removal would therefore increase overlap in unit recruitment across time and across contexts, reducing the separation of context-dependent subnetworks.

Together, these findings indicate that temporal dynamics are not sufficient in isolation, but become effective when coupled with sparse recruitment. Sparse activity partitions representations into partially context-specific subnetworks, whereas temporal dynamics further separate them across time; in particular, reset may reduce repeated co-recruitment of the same units, thereby limiting overlap and interference between contexts. This interpretation is consistent with our analytical framework (see Methods), in which threshold-induced sparsity constrains gradient updates to context-relevant subnetworks, while temporal dynamics further shape learning through state-dependent modulation.
\par\vspace{0.5\baselineskip}
\noindent\textbf{Sparse spatiotemporal spiking supports efficient lifelong learning under resource constraints}\\
Beyond accuracy, spiking computation is widely associated with energy efficiency because information is conveyed by sparse, event-driven spikes \cite{pei2019towards, gonzalez2024spinnaker2}. Consistent with this expectation, power estimation showed that SNNs consumed substantially less power than capacity-matched ANNs, with a 32.9\% reduction in total power (Fig.6a).

When SNNs were trained using biologically inspired local plasticity rules that do not require full error backpropagation, a distinct trade-off between accuracy and efficiency emerged. Using a SoftHebb plasticity rule \cite{journe2022hebbian}, SNNs exhibited a modest reduction in decoding accuracy compared with BP-trained SNNs, but consistently achieved performance that was comparable to or higher than ANN baselines across CIFAR-10, CIFAR-100, MNIST-Family, and T-ImageNet in class-incremental learning settings (Fig.6b). Importantly, this slight accuracy reduction was accompanied by substantial gains in computational efficiency. Relative to BP-trained SNNs, SoftHebb-trained SNNs showed a markedly reduced memory footprint and computation cost across batch sizes, with the estimated memory footprint decreasing from 11.2 MB to 7.1 MB and computation cost from 1.37 to 0.91 GFLOPs at the largest batch size tested (Fig.6c, 6d). Together, these results demonstrate that local plasticity enables SNNs to maintain competitive continuous learning performance while substantially reducing memory and computational demands.

Collectively, sparse spatiotemporal spike-based computation supports competitive lifelong-learning performance with reduced power and resource demands, suggesting a biologically grounded framework for scalable lifelong learning in artificial intelligence.

\section*{Discussion}

In this study, we identify two cooperative coding principles—sparse recruitment and temporal dynamics—that together support context-dependent behavioral switching across biological and artificial systems. 

In mouse mPFC, population activity is organized into both context-selective ensembles and a shared representational subspace: the selective component enables reliable separation of context-specific rules or policies, whereas the shared component preserves features that generalize across contexts (e.g., common stimulus attributes) \cite{Spanne_Jorntell_2015, tissot2025sensorimotor}. Our data further indicate that context decoding depends on intrinsic continuous-time dynamics, as decoding accuracy increases with the number of sampled time points and is reduced when temporal continuity or ordering is disrupted. These findings suggest that temporal structure contributes to context discriminability beyond the mere accumulation of activity over time.

Building on our observation that context reconfiguration in the mPFC relies on sparse recruitment and intrinsic temporal dynamics, we reasoned that the same spatiotemporal principles should be particularly beneficial in artificial settings where contexts change frequently and learning must proceed under nonstationary conditions—circumstances naturally captured by lifelong learning \cite{kudithipudi2022biological}. Consistent with this prediction, models (SNNs) that instantiate threshold-driven sparsity together with an internal temporal state exhibit improved lifelong-learning performance relative to matched static baselines, indicating that joint sparse coding and temporal dynamics are advantageous when knowledge must be updated without excessive interference with prior representations.

Importantly, our manipulations further indicate that the observed benefit is not a generic byproduct of simply adding a time axis or sparsifying a single processing stage. Rather, it reflects distinct but cooperative contributions from sparsity and temporal dynamics, with a critical dependence on the scope of sparsification. Sparsity primarily constrains cross-context interference by partitioning computation into context-preferential pathways, whereas temporal dynamics are not effective in isolation but, when coupled with sparse recruitment, further increase the separation of context-dependent activity across time (see Theoretical Analysis in Methods). In line with this complementarity, sparsification confined to a local subset of the network fails to isolate context-specific computation, as downstream representations remain densely shared. By contrast, network-wide sparsity promotes the formation of functional subnetworks that restrict credit assignment to context-relevant pathways and reduce representational overlap across tasks. Together, these findings argue that sparse recruitment must be sufficiently widespread to achieve effective context separation, and that intrinsic temporal dynamics become beneficial when coupled to this sparse organization, further reducing overlap and interference across contexts.

At the same time, the correspondence between neural data and computational models in our study is established at the level of shared population-level principles rather than detailed circuit matching. Biological systems implement sparse recruitment and temporal dynamics through specific microcircuit motifs, cell-type–dependent properties, and synaptic mechanisms. Bridging this gap will require tighter anatomical and cell-type–specific constraints, as well as causal perturbations targeting candidate timescale-generating mechanisms. Furthermore, although our models incorporate intrinsic temporal state, they implement a relatively restricted range of timescales. In contrast, cortical systems exhibit hierarchically organized and interacting temporal gradients shaped by recurrent circuitry and synaptic processes \cite{murray2014hierarchy, abbott2004synaptic, hasson2008hierarchy}. Extending artificial architectures to capture such multi-timescale structure may further enhance their ability to balance stability and flexibility in dynamic environments.

A common critique of spiking neural networks is that they lag behind conventional ANNs on static, i.i.d. benchmarks—where performance is largely determined by rapid fitting to a fixed data distribution—despite their well-recognized advantages in energy efficiency \cite{rueckauer2017conversion, yamazaki2022spiking, Maass1997, pei2019towards, shi2025hybrid}. Our results suggest that this apparent limitation reflects, at least in part, a mismatch between benchmark regimes and the computational strengths of spiking models. In natural settings, animals learn under dynamic, non-stationary conditions: environments change, contexts switch, and new information must be incorporated without erasing prior knowledge \cite{kudithipudi2022biological}. When these demands are formalized within a lifelong-learning framework, SNNs can exhibit a selective advantage, consistent with the view that event-driven sparsity and intrinsic temporal state are not optimized for one-shot static fitting, but for stable updating over time. This perspective offers a principled explanation for why biological neural systems rely on sparse recruitment and temporal dynamics to support flexible behavior while preserving memory in changing environments.

Beyond mechanistic insight, our work also highlights an efficiency-oriented learning direction that aligns with biological constraints. Pairing sparse spatiotemporal organization with bio-inspired learning rules—specifically SoftHebb—yields a learning paradigm that departs from backpropagation-based optimization while retaining strong performance under lifelong-learning conditions. More broadly, these results suggest that architectural principles inspired by neural population coding can support context-dependent computation through intrinsic network dynamics, reducing reliance on task-specific constraints. Together, this framework offers a scalable and energy-efficient route toward lifelong learning in artificial systems.

Finally, while sparse coding and temporal dynamics have been documented for decades in neural systems, our study advances these classical principles by demonstrating how their interaction supports flexible context configuration and lifelong learning, and by validating this role across both biological circuits and artificial lifelong-learning paradigms. By linking population-level coding strategies in the brain to performance advantages in dynamic learning regimes, our work helps bridge a longstanding gap between neural computation and artificial intelligence.

\newpage

\subsection*{Table 1. Comparison of properties between spiking neural networks and artificial neural networks.}

\begin{table}[h]
    \centering
    \renewcommand{\arraystretch}{1.3} % 增加行间距
    \begin{tabularx}{\textwidth}{lX X}
        \toprule
        \textbf{Properties} & \textbf{SNNs} & \textbf{ANNs} \\
        \midrule
        Value representations & Binary spike events (1/0) & Continuous values ([0,1], unbounded) \\
        Information encoding  & Sparse: selective neuron activation through threshold-mediated firing & Dense: all neurons are typically active \\
        Temporal dynamics & Dynamic: time-driven spike activity, encoding temporal information & Static: single-step processing, lacks intrinsic temporal representation \\
        Learning strategies & Unsupervised plasticity/BP & BP \\
        \bottomrule
    \end{tabularx}
\end{table}

\newpage

%% Main Figures
\noindent\includegraphics[width=0.95\linewidth]{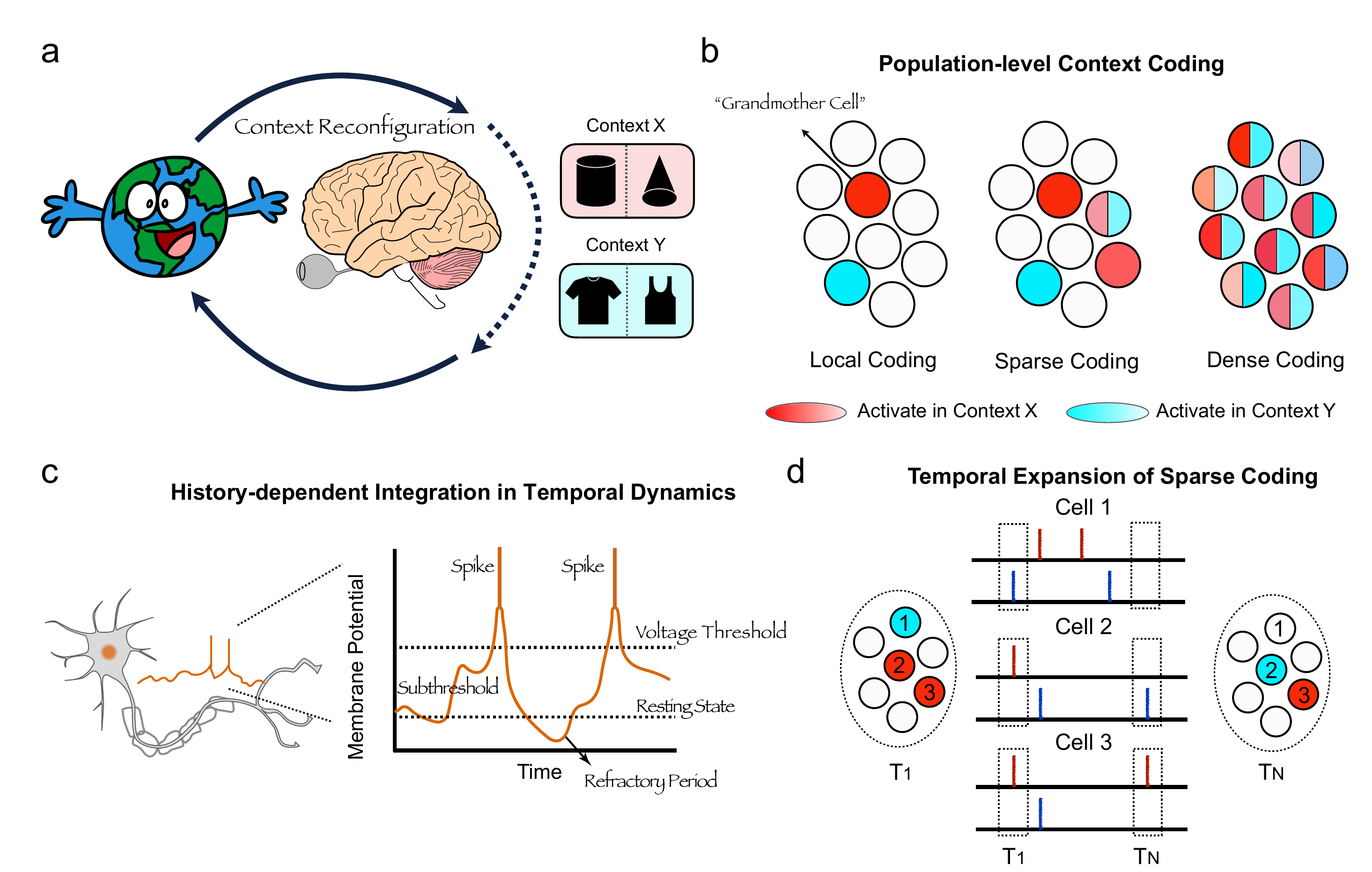}

\subsection*{Fig.1: Sparse spatiotemporal coding enables context reconfiguration in the brain.}

(a) Conceptual schematic of the brain interacting with a dynamic environment across shifting contexts. 
\newline
(b) Illustration of alternative coding schemes during a context shift between context X and context Y. Neurons active in context X are shown in red, neurons active in context Y are shown in blue, and bicolored nodes indicate neurons recruited in both contexts. 
\newline
(c) Schematic illustration of intrinsic temporal dynamics at the single-neuron level. Neuronal membrane potentials integrate synaptic inputs over time, giving rise to temporally extended firing-rate responses rather than instantaneous activation. Subthreshold integration and recurrent dynamics enable activity to accumulate across successive time points, providing a temporal substrate for stabilizing population-level representations during context transitions.
\newline
(d) Example firing-rate profiles of a single neuron over time under Context X and Context Y. Although mean firing rates are comparable across contexts, differences in the temporal structure of firing enable further discrimination between contexts when neural activity is considered over time.

\newpage

\noindent\includegraphics[width=0.95\linewidth]{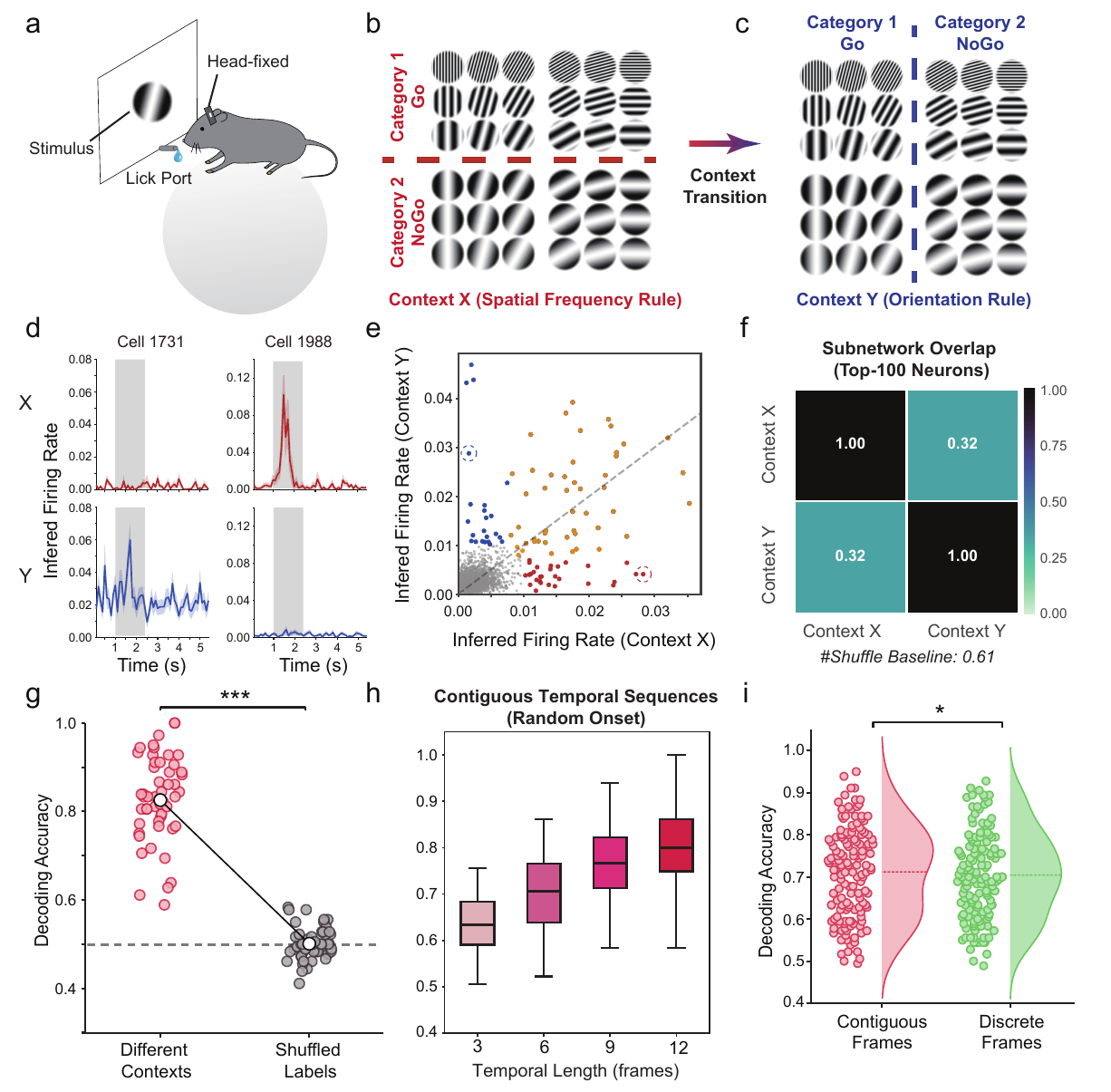}

\subsection*{Fig.2: Context representations in mPFC arise from ensemble-specific recruitment and temporal dynamics.}

(a) Schematic of the head-fixed behavioral setup. A mouse categorizes visual grating stimuli under a task rule that can switch across contexts. 
\newline
(b and c) The full stimulus set (36 gratings; 6 spatial frequencies × 6 orientations) presented under two contexts. In context X, stimuli are assigned to Category 1 or Category 2 based on spatial frequency (horizontal dashed boundary; B). After a context transition, the same stimuli are categorized in context Y based on orientation (vertical dashed boundary; C). 
\newline
(d) Trial-aligned inferred firing-rate traces for two example neurons (circled in E) during Context X (top) and Context Y (bottom). Shading/error bars indicate mean ± s.e.m. across trials. Gray bar denotes the stimulus presentation period.
\newline
(e) Neuron-wise comparison of inferred firing rates across contexts. Each dot represents one neuron. Colors indicate neurons ranked among the top 100 in both contexts (yellow), in context X only (red), in context Y only (green), or in neither set (gray). Dashed line indicates the identity line (y = x).
\newline
(f) Heat map showing the overlap fraction between the top 100 neurons ranked by inferred firing rate in context X and context Y. The off-diagonal value (0.32) indicates partial overlap between highly active neuronal populations across contexts and is markedly lower than a shuffle-derived chance level (0.61 $\pm$ 0.04), estimated by randomly permuting trial identities between contexts.
\newline
(g) SVM decoding accuracy for discriminating context (rule identity) from population activity, compared with shuffled-label controls. Each dot represents one decoding run using a randomly subsampled neuronal population and a random initialization. Black dots connected by line indicate paired comparisons across identical subsamples. The dashed line denotes chance level. ***p $<$ 0.001.
\newline
(h) Dependence of context decoding accuracy on temporal integration. Linear SVM decoding accuracy is shown as a function of temporal window length, with longer temporal integration yielding progressively higher decoding performance. Significant differences were observed between adjacent temporal windows.
\newline
(i) Comparison of decoding accuracy between contiguously sampled temporal sequences and discretely sampled temporal sequences. Each dot represents one decoding run using a randomly subsampled neuronal population and a random initialization; for visualization clarity, only a random subset of runs is shown. Kernel density estimates illustrate the distribution of decoding accuracies computed from all runs. Decoding was performed using temporal windows containing 2–13 frames; window lengths of 1 frame and the full sequence (14 frames) were excluded, as contiguously and discretely sampled sequences are equivalent under these conditions. Contiguous temporal sequences yielded significantly higher decoding accuracy than discretely sampled temporal sequences (*p $=$ 0.0426).
\newline
Unless otherwise stated, biological analyses in d--i used response-matched trials from 2,306 neurons recorded from 10 mice.

\newpage

\noindent\includegraphics[width=0.95\linewidth]{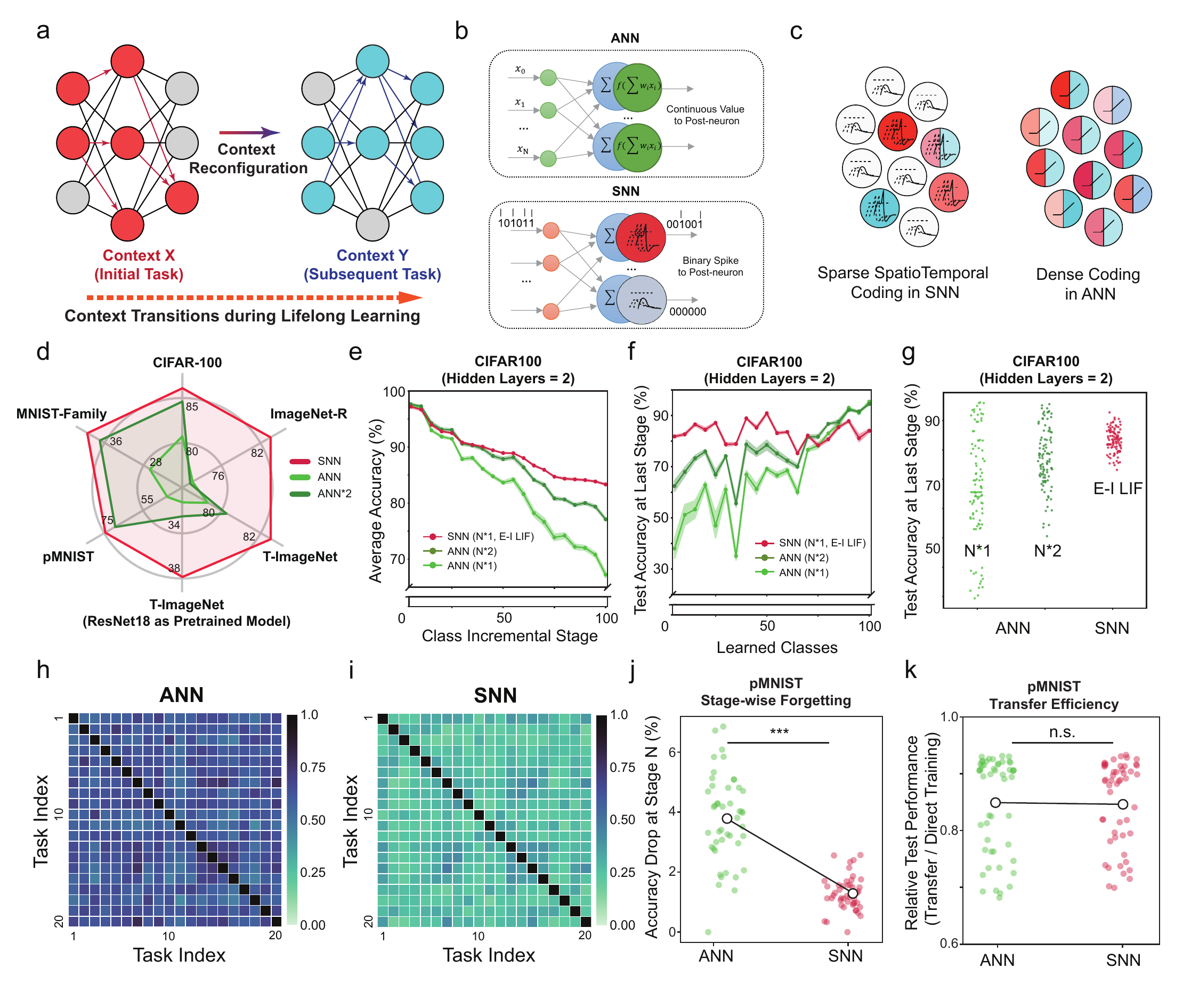}

\subsection*{Fig.3: Built-in threshold-induced sparsity and temporal coding enable superior lifelong learning in SNNs.}

(a) Schematic illustration of context-dependent reconfiguration during lifelong learning. A shared neural network undergoes functional reconfiguration following a context transition, supporting an initial task (Context X) and a subsequent task (Context Y) while preserving the same underlying architecture.
\newline
(b) Architectural comparison between artificial neural networks and spiking neural networks. ANNs rely on continuous-valued representations, whereas SNNs communicate via binary spikes with internal temporal dynamics.
\newline
(c) Conceptual comparison of coding schemes under a context shift between context X and context Y, highlighting sparse spatiotemporal coding in SNNs versus denser coding in ANNs. Neurons preferentially active in context X are shown in red, neurons preferentially active in context Y are shown in blue, and bicolored nodes indicate neurons recruited in both contexts.
\newline
(d) Radar plot comparing final-stage accuracy across multiple lifelong-learning benchmarks (CIFAR-100, ImageNet-R, T-ImageNet, pMNIST, and MNIST-Family).
\newline
(e) Average test accuracy across class-incremental stages on CIFAR-100 (mean $\pm$ SEM across five random seeds).
\newline
(f) CIFAR-100 class-incremental learning (two hidden layers). Final-stage test accuracy (after completing all incremental stages) is plotted for ANNs with matched capacity (N×1) or doubled capacity (N×2) and SNN (N×1, E-I LIF). Curves show mean $\pm$ SEM across five random seeds.
\newline
(g) Distribution of final-stage test accuracy across all learned classes on CIFAR-100. Each dot denotes one task from one random seed. 
\newline
(h and i) Task-by-task overlap of high-activity subnetworks. Heat maps show the overlap fraction of the top 100 neurons ranked by firing rate between all pairs of tasks for (H) ANN and (I) SNN.
\newline
(j) Between-group forgetting, quantified as the accuracy drop when transitioning from one task group to the next (computed per random seed). Each dot denotes one group transition from one seed; ***p $<$ 0.001.
\newline
(k) Transfer efficiency on pMNIST, defined as the ratio of performance with direct training on the target task to performance after training on other similar tasks and testing on the target task; n.s. No Significance.

\newpage

\noindent\includegraphics[width=0.95\linewidth]{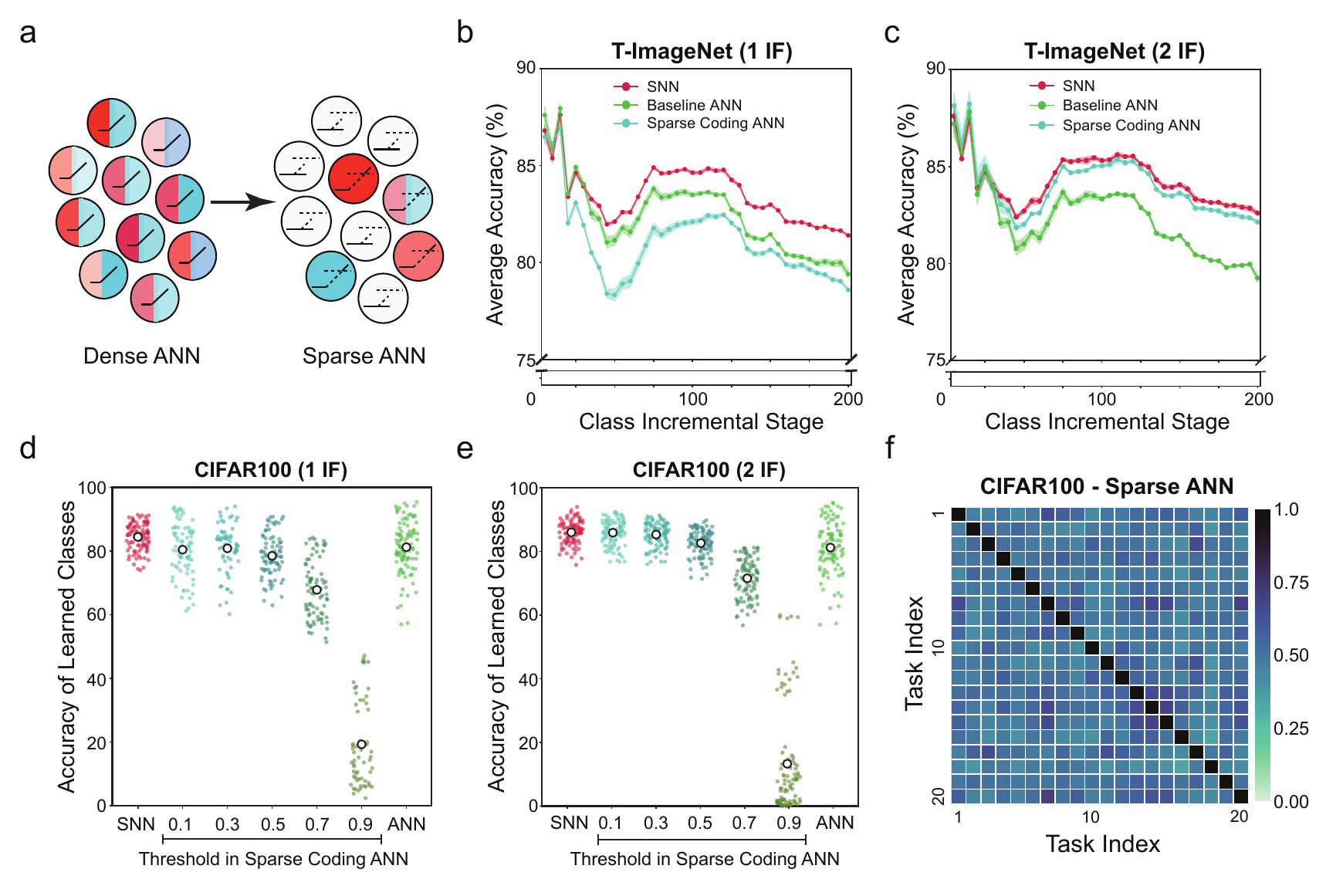}

\subsection*{Fig.4: Network-wide sparsity improves lifelong learning performance.}

(a) Schematic illustration of the transition from dense to sparse coding in ANNs. Neurons preferentially active in context X are shown in red, and those preferentially active in context Y are shown in blue. Bicolored nodes indicate neurons recruited in both contexts, whereas uncolored nodes denote neurons that remain inactive under the threshold constraint.
\newline
(b, c) Average test accuracy across class-incremental stages on (b) T-ImageNet (1 IF) and (c) T-ImageNet (2 IF), which are shown as mean $\pm$ SEM across five random seeds.
\newline
(d, e) Distribution of final-stage test accuracy across all learned classes for various sparsity levels on (d) CIFAR-100 (1 IF) and (e) CIFAR-100 (2 IF). Each dot denotes one run (random seed); sparsity thresholds (th) are indicated.
\newline
(f) Task-by-task overlap of high-activity subnetworks. Heat maps show the overlap fraction of the top 100 neurons ranked by firing rate between all pairs of tasks for sparse ANN.

\newpage

\noindent\includegraphics[width=0.95\linewidth]{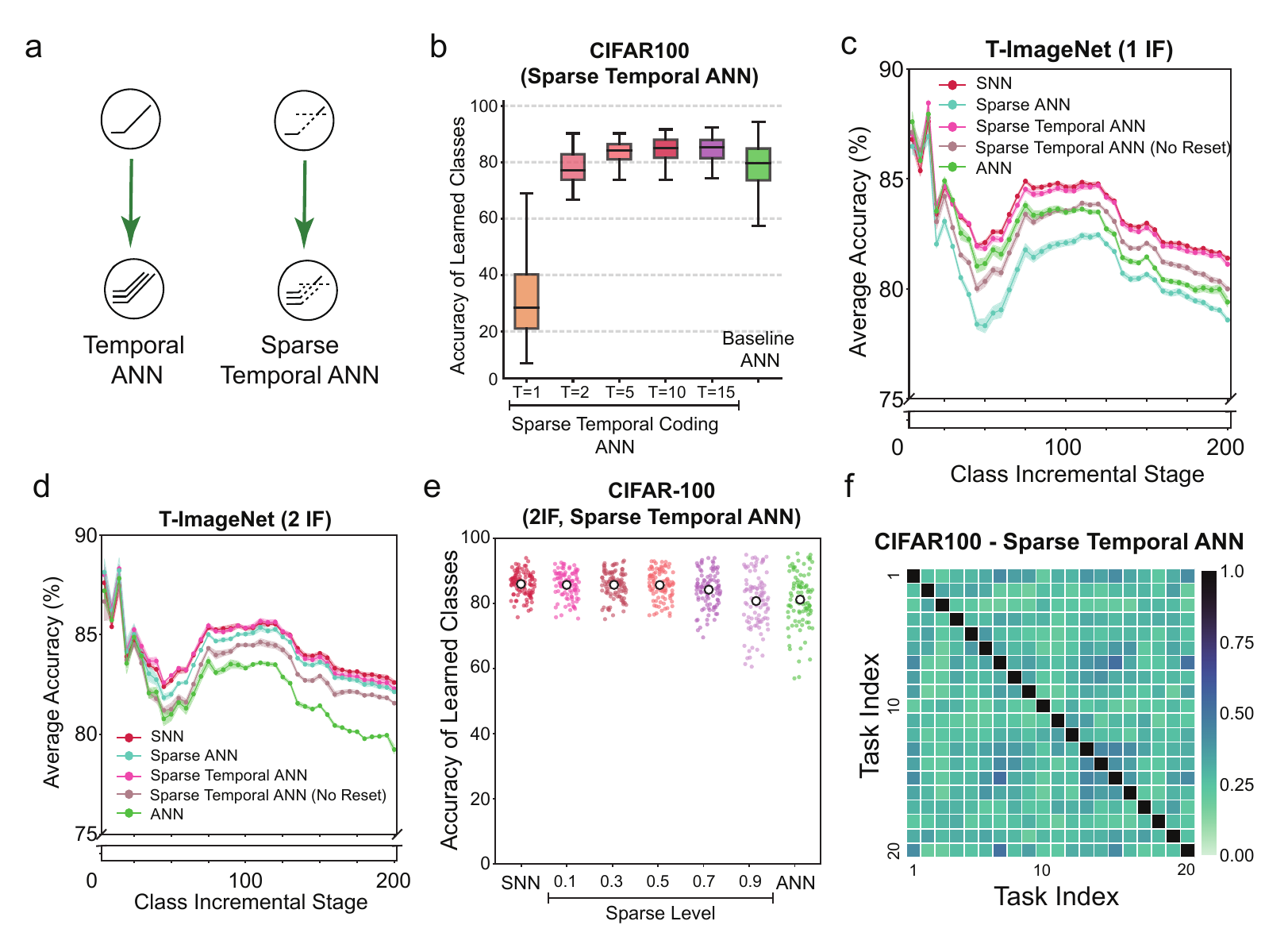}

\subsection*{Fig.5: Temporal dynamics interact with sparsity to enhance context separation and improve lifelong learning.}

(a) Schematic illustrating the introduction of internal temporal dynamics into an ANN.
\newline
(b) Box-and-whisker plots of last-stage average accuracy across random seeds across increasing time length (T) for sparse temporal coding ANNs. Center lines indicate the median, boxes indicate the interquartile range (IQR), and whiskers indicate the range (minimum to maximum).
\newline
(c, d) Mean test accuracy ($\pm$ SEM across five seeds) across class-incremental stages on (c) T-ImageNet (1 IF) and (d) T-ImageNet (2 IF).
\newline
(e) Distribution of final-stage test accuracy across all learned classes for various sparsity level on sparse temporal coding ANN for CIFAR-100 (2 IF). Each dot denotes one run (random seed); sparsity thresholds (th) are indicated.
\newline
(f) Task-by-task overlap of high-activity subnetworks. Heat maps show the overlap fraction of the top 100 neurons ranked by firing rate between all pairs of tasks for sparse temporal coding ANNs.

\newpage

\begin{center}
 \noindent\includegraphics[width=0.8\linewidth]{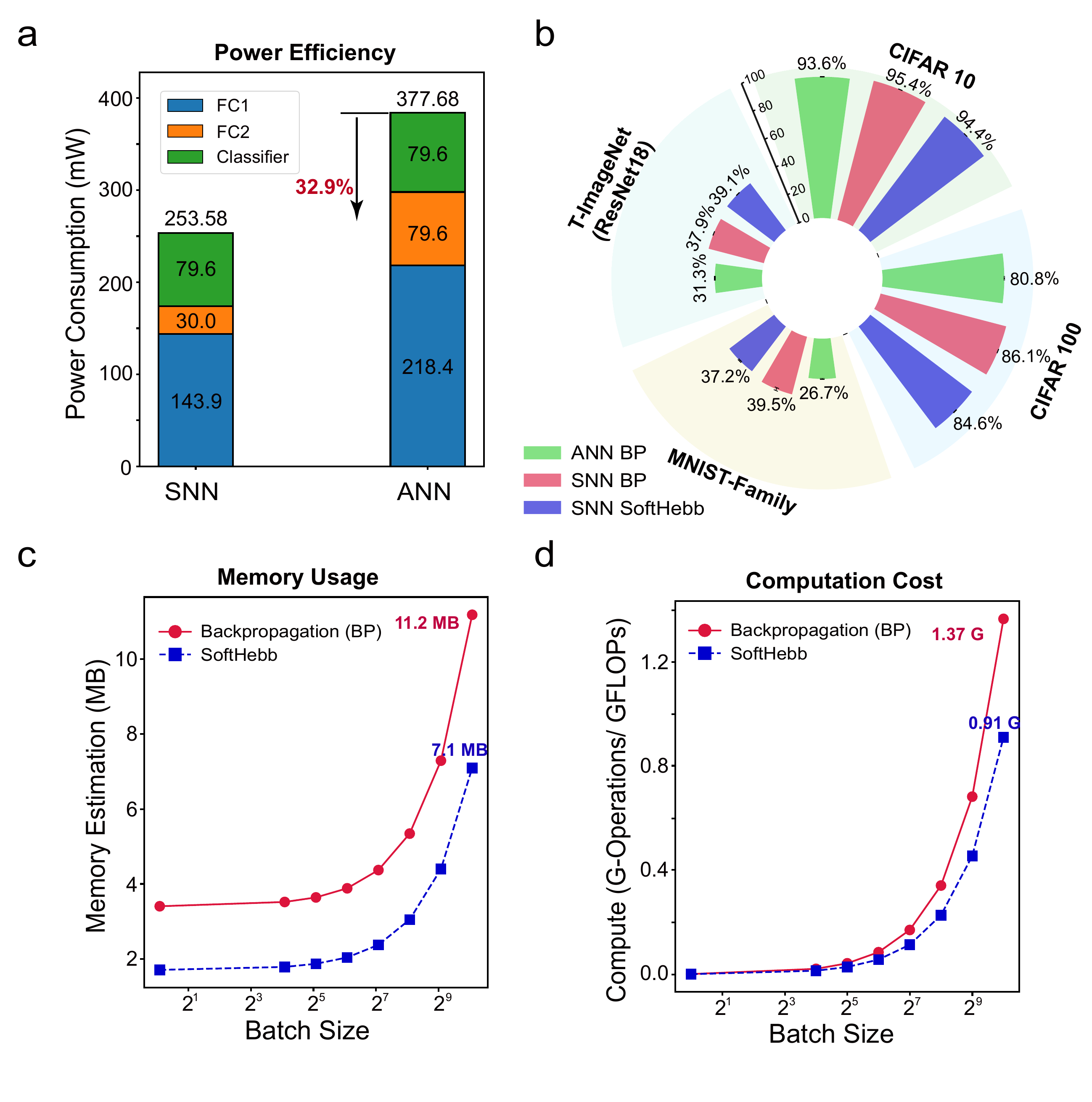}   
\end{center}

\subsection*{Fig.6: SNNs support scalable lifelong learning with low power and reduced resource demands.}

(a) Power consumption comparison between SNNs and ANNs, decomposed into contributions from the first fully connected layer (FC1), second fully connected layer (FC2), and the classifier.
\newline
(b) Radial bar plot showing classification accuracy (\%) of ANN trained with backpropagation (ANN BP), SNN trained with backpropagation (SNN BP), and SNN trained with SoftHebb learning (SNN SoftHebb) across CIFAR-10, CIFAR-100, MNIST-Family, and TinyImageNet (ResNet-18 pretrained). Error bars indicate SEM.
\newline
(c and d) Estimated (c) memory footprint and (d) computation cost for backpropagation and SoftHebb across batch sizes; values at the largest batch size are indicated.

\newpage

\bibliographystyle{unsrtnat}
\bibliography{bibliography}

\newpage

\section*{Methods}
\subsection*{Neural experiment details}
\bmhead{Animals}
We analyzed mPFC population recordings from adult female C57BL/6 mice engaged in a rule-based Go/NoGo visual categorization task, using data from a previously published study \cite{reinert2021mouse}. The present reanalysis included recordings from 10 mice (M01--M05 and M07--M11). Mouse M06 was excluded from the present analysis because matched data from both contexts were not available. All animal procedures were performed and approved as described in the original publication.
\bmhead{Visual stimuli and behavioral task}
Visual stimuli consisted of drifting sinusoidal gratings defined by combinations of orientation and spatial frequency and were presented for $\text{~}$1.3s per trial. A categorization rule assigned stimuli to Go or NoGo categories based on either orientation (boundary at $45^\circ$) or spatial frequency (boundary at 0.043 cycles/$^{\circ}$). Mice were first trained on one rule and subsequently retrained on the alternative rule using the same stimulus set, yielding two task contexts. In this study, analyses focused on two task stages: rule~1 (context~X, spatial-frequency discrimination) and rule~2 (context~Y, orientation discrimination), which shared matched stimulus sets across contexts.
\bmhead{Two-photon imaging}
Two-photon calcium imaging data from the mPFC were obtained from a previously published dataset \cite{reinert2021mouse}, in which neuronal activity was recorded using the genetically encoded calcium indicator GCaMP6m, providing a characteristic temporal resolution for the measured population activity. Imaging data were acquired at 10~Hz and synchronized with stimulus presentation and behavioral events in the original study.

In the present work, we analyzed frame-wise inferred neuronal activity from sessions T5 and T8. Neural activity was aligned to stimulus onset and analyzed separately under different contextual conditions.
\bmhead{Data preprocessing}
Across datasets from contexts X and Y, a total of 2{,}306 neurons from 10 mice and 15 imaging regions were included for analysis. Specifically, one imaging region was included for each of M01–M05, and two imaging regions were included for each of M07–M11. A randomly assigned subset of mice were trained first on context X and second on context Y, and another subset of animals was trained first on context Y, and then context X.

To ensure context-specific analyses and minimize potential confounds from recording order, data were aligned by context rather than session chronology. Trials containing missing values (NaNs) in neuronal activity traces were excluded, resulting in the removal of 10 trials in total. Details of the excluded trials are provided in Supplementary Table~S1.

To control for Go/NoGo-related motor responses, we used only response-matched trials across contexts, including stimuli that were Go in both contexts (orientation $>$ $45^\circ$ and sf $>$ 0.043 cycles/$^{\circ}$) and stimuli that were NoGo in both contexts (orientation $<$ $45^\circ$ and sf $<$ 0.043 cycles/$^{\circ}$). To enable concatenation of neuronal activity across animals, the number of trials was then matched across mice by subsampling. For each context, the minimum number of available trials was used, with trials randomly sampled without replacement for animals with larger trial counts.
\bmhead{Ensemble overlap (top-100 analysis)}
Ensemble similarity across contexts was quantified using a top-$k$ overlap analysis based on inferred firing rates. For each neuron and each context, inferred firing rates were averaged across trials to obtain a context-specific mean firing rate. Neurons were ranked accordingly, and the top 100 neurons were selected to define the ensemble for each context.

Ensemble overlap between contexts $X$ and $Y$ was quantified using the Jaccard index, $|E_X \cap E_Y| / |E_X \cup E_Y|$, where $E_X$ and $E_Y$ denote the sets of top-100 neurons in the two contexts.
\bmhead{Context tuning index and neuron selection}
To quantify context selectivity at the population level, we defined a context tuning index (CxTI), conceptually related to the category tuning index reported previously \cite{reinert2021mouse} but adapted to specifically capture neural discrimination between contexts. For each animal, CxTI was computed for every recorded neuron using the mean inferred spike rate during stimulus presentation.

Specifically, based on the mean inferred spike rate during stimulus presentation, we quantified (i) across-context difference $D_{\text{across}}$: the average absolute difference in neuronal responses between stimuli belonging to different contexts and (ii) within-context difference $D_{\text{within}}$: the average absolute difference between stimuli within the same context. The across-context difference is identical for both contexts, whereas the within-context difference is context-dependent. These quantities were normalized by their sum to yield the CxTI. The CxTI for the \(i^{\mathrm{th}}\) neuron under context \(r\) was defined as:

\begin{equation}
CxTI^{(i)}_r = \frac{D_{\text{across}}^{(i)} - D_{\text{within}}^{(i)}}{D_{\text{across}}^{(i)} + D_{\text{within}}^{(i)} + \epsilon},
\end{equation}

\begin{equation}
    D_{\text{across}}^{(i)} = \frac{1}{N_1 N_2} \sum_{k=1}^{N_1} \sum_{l=1}^{N_2} \left| s_k^{(1)} - s_l^{(2)} \right|,
\end{equation}

\begin{equation} 
    D_{\text{within}}^{(i)} = \frac{1}{N^2_r} \sum_{k=1}^{N_r} \sum_{l=1}^{N_r} \left| s_k^{(1)} - s_l^{(1)} \right|, 
\end{equation}
where \(s_j^{(1)}\) and \(s_j^{(2)}\) denote the mean inferred spike rate of the \(i^{\mathrm{th}}\) neuron in response to the \(j^{\mathrm{th}}\) stimulus under the first and second contexts, respectively. \(N_r\) denotes the number of unique stimuli presented under context \(r\), and \(\epsilon\) is a small constant added for numerical stability. CxTI values range from $-1$ to $1$, with values close to $1$ indicating stronger context selectivity, whereas values near or below $0$ indicate weak or no context selectivity. For each context, we retained the top $10\%$ of neurons with the highest CxTI values and discarded the remaining neurons. The union of the selected neurons from both contexts was then used as the feature set for subsequent classification.
\bmhead{SVM decoding}
Using CxTI-selected neurons, SVM decoding was performed to classify context from neuronal spike rates. A time series–specific SVM was adopted to accommodate the temporal structure of the sampled neuronal responses, allowing each trial to be treated as a temporal sequence rather than a vectorized feature representation. Neurons from all mice were concatenated, and for each analysis, $32$ neurons were randomly sampled. For each sampled neuron set, spike-rate activity was extracted over a fixed post-stimulus period (frames $11$–$24$ out of 54 imaging frames, following \cite{reinert2021mouse}). Both neuron sampling and decoding were repeated ten times using five different random seeds. For each sample, $80\%$ of the data were used for training and the remaining $20\%$ for testing. Features were standardized using z-score normalization based on the training set. SVM hyperparameters were optimized using grid search with 10-fold cross-validation on the training data. To assess the specificity of CxTI-based neuron selection, the decoding analysis was repeated using randomly selected neurons under the same sampling and training procedures. In addition, to control for potential confounding factors, the entire analysis was repeated after randomly shuffling the context labels.
\bmhead{Temporal integration analysis}
Using the same CxTI-based decoding pipeline, we examined how temporal aggregation affected context decoding. Rather than using all imaging frames within the post-stimulus window, subsets of time points were sampled to vary the amount of temporal information and whether temporal continuity was preserved. For each trial, neuronal responses were sampled at $n$ time points ($n = 2$--$14$) from frames $11$--$24$ using two temporal sampling strategies: (i) temporally unconstrained random sampling of $n$ time points, or (ii) random sampling of a temporally contiguous segment of length $n$. All remaining settings were identical to the CxTI-based decoding analysis. The procedure was repeated using five different random seeds, with ten independent repetitions of neuron sampling and temporal subsampling.

To quantify the effect of temporal continuity, decoding accuracies obtained with temporally contiguous sampling were compared with those obtained using temporally unconstrained sampling. Accuracies were pooled across repetitions and values of $n$ included in the analysis ($n = 3$--$12$). Values of $n = 2$ and $n = 13$ were excluded because temporal continuity provides limited differentiation at very small or near-complete temporal coverage, and $n = 14$ was excluded because the two sampling strategies are identical when all available time points are used.

To further isolate the contribution of temporal ordering beyond continuity, we introduced an additional control condition based on the temporally contiguous samples. Specifically, for each trial, the frames obtained from the temporally contiguous segment were randomly permuted, thereby preserving the set of time points while disrupting their temporal order. Decoding accuracies obtained with the original (ordered) contiguous sequences were then compared with those obtained from the shuffled sequences. This analysis was restricted to $n = 13$ and $n = 14$, where a larger number of time points provides richer temporal information, such that shuffling more effectively disrupts sequential structure and thus more sensitively reflects the contribution of temporal ordering.
\subsection*{Computational modeling and corresponding analysis}
\bmhead{Lifelong-learning benchmarks and protocols}
We evaluated models under three lifelong-learning scenarios: class-incremental learning (CIL), task-incremental learning (TIL), and domain-incremental learning (DIL). Learning proceeded in discrete stages, where new classes or tasks were introduced sequentially and data from previous stages were not revisited.

In CIL, each stage introduced $n$ novel classes, yielding $N=nk$ classes after $k$ stages, and inference was performed over all previously learned classes. We implemented CIL using a fixed-size classifier initialized with $N$ outputs and a stage-wise class mask during training. Specifically, for stage $i\in\{1,\dots,k\}$, given logits $\mathbf{z}\in\mathbb{R}^{N}$, a binary mask $\mathbf{m}^{(i)}\in\{0,1\}^{N}$ was applied to obtain masked logits $\tilde{\mathbf{z}}^{(i)} = \mathbf{m}^{(i)} \odot \mathbf{z}$, where $m^{(i)}_{c}=1$ if $(i-1)n < c \le in$ and $0$ otherwise. The training loss was computed only on the active subset defined by $\tilde{\mathbf{z}}^{(i)}$.

In TIL and DIL, each stage corresponded to a new $n$-way classification task. Performance was evaluated per task. Task identity was provided at inference in TIL (to select the corresponding output head or decision rule), but not provided in DIL, requiring task-agnostic prediction with an $n$-way classifier.

For each lifelong learning experiment, all models were evaluated using five independent random seeds to ensure robustness of the reported results.
\bmhead{Dataset}
For the permuted MNIST (pMNIST) benchmark, we constructed 40 tasks by applying fixed pixel permutations to the original MNIST images, which were organized into 10 groups of four with higher within-group similarity.

MNIST, Fashion-MNIST, and KMNIST were concatenated to form a 30-class dataset referred to as \emph{MNIST-Family}, which was partitioned into 15 tasks with two classes per task.

CIFAR-10, CIFAR-100, Tiny-ImageNet, and ImageNet-R were partitioned into 5, 20, and 40 episodes, respectively, with two classes per episode for CIFAR-10 and five classes per episode for the remaining datasets. All $32 \times 32$ RGB images were encoded using a pretrained CLIP vision encoder to obtain 768-dimensional feature representations; for Tiny-ImageNet, features were additionally extracted using a pretrained ResNet-18.
\bmhead{Network architectures}
Unless otherwise noted, ANNs consisted of one or two fully connected (FC) hidden layers, each followed by ReLU and Batch Normalization (BN), and a final linear classification layer.

To ensure architectural comparability, SNNs were constructed by temporally unfolding the ANN architecture over $T$ time steps and introducing spiking neuron dynamics before the FC layers, while removing ReLU activations. All other architectural components, including layer dimensionality and normalization, were kept identical.

Static inputs were replicated across time:
\begin{equation}
\mathbf{x}_{t}= \mathbf{x}, \quad t=1,\dots,T .
\label{eq:repeat_input}
\end{equation}
Spiking neuron dynamics layers in SNNs were implemented using either leaky integrate-and-fire (LIF) neurons or ternary leaky integrate-and-fire (TLIF) neurons. For both models, membrane potential evolved according to
\begin{equation}
u_t = \tau u_{t-1} + I_t ,
\label{eq:mem_update}
\end{equation}
where $\tau\in(0,1)$ is the leak factor and $I_t$ denotes the synaptic input.

For LIF neurons, spike emission followed
\begin{equation}
s_t = H\!\left(u_t - v_{\mathrm{th}}\right),
\end{equation}
where $s_t\in\{0,1\}$ and $H(\cdot)$ is the Heaviside step function.

For TLIF neurons, spike emission was ternary:
\begin{equation}
s_t =
\begin{cases}
1, & u_t \ge v_{\mathrm{th}},\\
-1, & u_t \le -v_{\mathrm{th}},\\
0, & \text{otherwise},
\end{cases}
\end{equation}
corresponding to excitatory ($+1$), inhibitory ($-1$), and silent ($0$) states. After spike emission, the membrane potential was reset as:
\begin{align}
    u_t &\leftarrow (1-s_t)\,u_t.
\end{align}
We evaluated two SNN configurations that differed in the depth at which spiking dynamics were applied. In the first configuration, spiking dynamics were applied only before the hidden fully connected (FC) layer, whereas in the second configuration spiking dynamics were applied both before the hidden FC layer(s) and before the classifier layer.

Given input spikes $\mathbf{s}_t$, shared hidden-layer computations were
\begin{align}
\mathbf{z}_t &= W_1 \mathbf{s}_t + \mathbf{b}_1, \\
\mathbf{z}_t^{(1)} &= \phi(\mathbf{z}_t),
\end{align}
where $\phi(\cdot)$ denotes Batch Normalization (BN).

In the first configuration, the classifier operated directly on the continuous hidden activation,
\begin{equation}
\mathbf{o}_t = W_2 \mathbf{z}_t^{(1)} + \mathbf{b}_2 .
\label{eq:First_config_output}
\end{equation}
In the second configuration, hidden activations were first transformed into spikes by a spiking neuron module $IF(\cdot)$ (LIF or TLIF),
\begin{align}
\mathbf{s}_t^{(1)} &= IF(\mathbf{z}_t^{(1)}), \\
\mathbf{o}_t &= W_2 \mathbf{s}_t^{(1)} + \mathbf{b}_2 .
\end{align}
Here, $\mathbf{z}_t^{(1)}$ was treated as the synaptic input current to the spiking neuron module.

Unless otherwise noted, SNN outputs were decoded using rate coding by averaging classifier logits over time,
$\mathbf{o} = \frac{1}{T}\sum_{t=1}^{T}\mathbf{o}_t$.
\bmhead{Sparsity mechanisms introduced to ANNs}
To introduce sparsity in ANNs, we replaced ReLU with a thresholded activation applied before FC layers. Given the pre-activation $h$, a bounded nonlinearity $r(\cdot)$ (e.g., $\tanh$) was first applied, followed by a magnitude threshold $\mathrm{th}$:
\begin{equation}
h^{(1)} =
\begin{cases}
r(h), & |r(h)| > \mathrm{th}, \\
0, & \text{otherwise}.
\end{cases}
\label{eq:sparse_activation}
\end{equation}
Because the hard threshold is non-differentiable, we used a surrogate-gradient approximation during backpropagation. Specifically, the gradient through the gating operation was approximated by the derivative of a sigmoid function,
\begin{equation}
\frac{\partial h^{(1)}}{\partial h} \approx r'(h)\,\sigma(\alpha h)\bigl(1-\sigma(\alpha h)\bigr),
\label{eq:surrogate_grad}
\end{equation}
where $\sigma(x)=\frac{1}{1+e^{-x}}$ and $\alpha$ controls the surrogate steepness (set to $\alpha=1$ unless otherwise noted).
\bmhead{Temporal dynamics introduced to ANNs}
To introduce temporal dynamics in ANNs while keeping all operations continuous, we adopted LIF/TLIF-inspired membrane integration and reset mechanisms, but used continuous activations instead of spikes. Static inputs were replicated across time as in Eq.~(4). For each layer, the membrane potential was updated as Eq.~(\ref{eq:mem_update}).

The layer output was obtained by applying an activation operator $G(\cdot)$ to the membrane potential,
\begin{equation}
h_t = G(u_t).
\label{eq:ann_out}
\end{equation}
Depending on the experimental setting, $G(\cdot)$ was chosen as (i) an identity mapping ($G(u)=u$; no activation), (ii) a bounded nonlinearity ($G(u)=r(u)$, e.g., $\tanh$), or (iii) the thresholded sparse activation defined in Eq.~(\ref{eq:sparse_activation}) applied to $u_t$ (i.e., $h_t = h^{(1)}$ with $h \leftarrow u_t$).

A reset mechanism was applied analogously to spiking models. Specifically, the membrane potential was reset to zero whenever $h_t\neq 0$:
\begin{equation}
u_t \leftarrow
\begin{cases}
u_t, & h_t = 0, \\
0, & \text{otherwise}.
\end{cases}
\label{eq:reset_ann}
\end{equation}
\bmhead{SoftHebb learning}
SoftHebb learning \cite{journe2022hebbian} was used as a biologically plausible alternative to backpropagation for training hidden layers. Conventional hidden layers were replaced with SoftHebb linear layers while preserving the original weight dimensionality.

For each task, training proceeded in two phases. Upon task arrival, SoftHebb layers were updated in an unsupervised manner using forward propagation only, while all other network components were kept fixed. Synaptic weights in the SoftHebb layers were updated according to:
\begin{equation}
\Delta w_{ij}^{(SoftHebb)} = \eta \, y_j \left(x_i - u_j w_{ij} \right),
\label{eq:softhebb_update}
\end{equation}
where $x_i$ denotes the presynaptic activity, $u_j$ the total weighted pre-activation input of postsynaptic neuron $j$, $y_j$ the softmax-normalized output of neuron $j$, $w_{ij}$ the synaptic weight from neuron $i$ to neuron $j$, and $\eta$ the learning rate.

The classification head was subsequently trained using standard backpropagation with the SoftHebb layers fixed. Separate SGD optimizers were used for the SoftHebb layers and the classification head. In practice, the SoftHebb layers and the classification head were trained for one and five epochs per task, respectively.
\subsection*{Theoretical analysis}
\bmhead{Key parameters of SNNs in the backpropagation process}
For clarity, we consider the first SNN configuration with a single hidden layer (Eq.~(\ref{eq:First_config_output})) and a rate-coded readout.

We optimized the cross-entropy loss
\begin{align}
    L = CE(\overline{o}, y)=-\sum_{c=1}^{C} y_c \log(o_c)
\end{align}
where $o_c$ represents the predicted probability for each class, $y_c$ is the true label's probability distribution, and $C$ is the number of classes.\\
Assuming that the one-hot vector of the target label $y$ is denoted as $e_y$, the gradient of the loss with respect to the average output is:
\begin{align}
    \frac{\partial{L}}{\partial{\overline{o}}}
        =softmax(\overline{o})-e_y\overset{\triangle}{=}\overline{\delta}
\end{align}
Consequently, the gradient with respect to the instantaneous output $o_t$ at each time step is given by
\begin{align}
    \frac{\partial{L}}{\partial{o_t}}
        =\frac{1}{T}\overline{\delta}
\end{align}
When updating the weight matrix $W_1$ in the main layer of the network, the update rule can be expressed as:
\begin{align}
    \frac{\partial L}{\partial W_1}
        &= \sum_{t=1}^{T}\frac{1}{T}\frac{\partial L}{\partial z_t^{(1)}} s_t^{T} \nonumber\\
        &= \sum_{t=1}^{T} \alpha_t W_2 \overline{\delta}\,
           H(a_t-v_{th})
\end{align}
where $\alpha_t$ represents constant, notice the term $H(a_t - v_{th})$ constrains the update of specific neurons according to their spiking states.

The weight update rule for $\theta^{(n+1)}$ can then be written as:
\begin{align}
\theta^{(n+1)} &= \theta^{(n)} - \eta\sum_{t=1}^{T} \alpha_t W_2 \overline{\delta}\,
           H(a_t-v_{th})
\end{align}
where $\eta$ is the learning rate.

The incorporation of temporal dynamics (through $H(a_t - v_{th})$) introduces additional degrees of freedom (DOF) for task-dependent weight adaptation. This allows different neurons to be selectively updated based on their spiking conditions, providing more flexibility in learning and enabling the network to adapt to different task demands.
\subsection*{Resource efficiency}
\bmhead{Power estimation}
Power consumption of ANNs and SNNs was estimated using a hardware-level framework based on the Tianjic architecture \cite{pei2019towards}. All evaluations were conducted under identical hardware assumptions, including the same computing core, clock frequency, supply voltage, and technology node.
\bmhead{Memory footprint and computation cost estimation}
Because most neuromorphic platforms do not natively support both BP-based and feedforward-only learning within a unified architecture, learning-related power consumption was evaluated using a custom learning unit implemented in Verilog and based on the Tianjic computing core. Hardware synthesis and power estimation were performed using Synopsys Design Compiler with the UMC 28\,nm standard-cell library. Computational cost was quantified as the total number of floating-point operations (FLOPs) required during training and is reported in giga floating-point operations (GFLOPs).

As memory access and activation storage dominate energy consumption in neuromorphic systems, storage requirements were used as a proxy for relative power consumption. To compare resource demands of different learning rules, memory footprint was decomposed into static and dynamic components, corresponding to parameter storage and training-time activations, respectively.

For backpropagation, static storage accounted for both weights and gradients, and dynamic storage for batch-wise activations, yielding $S_{\mathrm{BP}}^{\mathrm{stat}} = 2Pb$ and $S_{\mathrm{BP}}^{\mathrm{dyn}} = B(H + O)b$. For SoftHebb learning, static storage included only synaptic weights and dynamic storage scaled with hidden-layer activations, yielding $S_{\mathrm{SH}}^{\mathrm{stat}} = Pb$ and $S_{\mathrm{SH}}^{\mathrm{dyn}} = BHb$. Here, $P$ denotes the number of trainable parameters, $B$ the batch size, $H$ the number of hidden units whose activations are stored during training, $O$ the number of output units, and $b$ the storage cost per parameter. Superscripts $\mathrm{stat}$ and $\mathrm{dyn}$ denote static and dynamic storage, respectively. Reduced dynamic storage under SoftHebb learning implies lower memory access demands and was used as an indicator of reduced power consumption.
\subsection*{Quantification and statistical analysis}
\bmhead{Transfer and Forgetting Metrics on pMNIST}
We used the pMNIST benchmark to quantify transfer efficiency and forgetting in a lifelong learning setting. Metrics were computed at the group level, where every four consecutive similar tasks were treated as one group.

\bmhead{Transfer efficiency}
For each group, transfer efficiency was evaluated on the fourth task using a two-stage protocol. After training on the first three tasks, the model was evaluated on the fourth task without prior training and then re-evaluated after explicit training. Transfer efficiency was defined as the ratio between the accuracy before training and the accuracy after training on the same task.

\bmhead{Forgetting rate}
To quantify forgetting across groups, we recorded the mean accuracy of each group immediately after its completion. After training the subsequent group, the previously completed group was re-evaluated, and the difference between its original and updated mean accuracy was defined as the forgetting rate. This procedure was repeated sequentially for all groups.
\bmhead{Statistics}
Statistical significance was assessed using nonparametric tests. For paired comparisons, a two-sided Wilcoxon signed-rank test was used. For unpaired comparisons, a two-sided Mann--Whitney U test was applied. No assumptions of normality were made. Statistical significance was denoted as follows: *$p < 0.05$, **$p < 0.01$, ***$p < 0.001$; n.s., not significant.

\section*{Author contributions}

L.S. and Q.S. conceptualized the study. Q.S., F.L., and L.S. developed the methodology. Q.S., Y.C., F.L., H.L., and M.X. conducted the investigation. S.R. and P.M.G. provided the biological data. Q.S., Y.C., and H.L. wrote the original draft. L.S., R.Z., S.R., and P.M.G. reviewed and edited the manuscript. L.S. supervised the project.

\section*{Competing interests}

The authors declare no competing interests.

\newpage

\begin{appendices}

\noindent\includegraphics[width=0.95\linewidth]{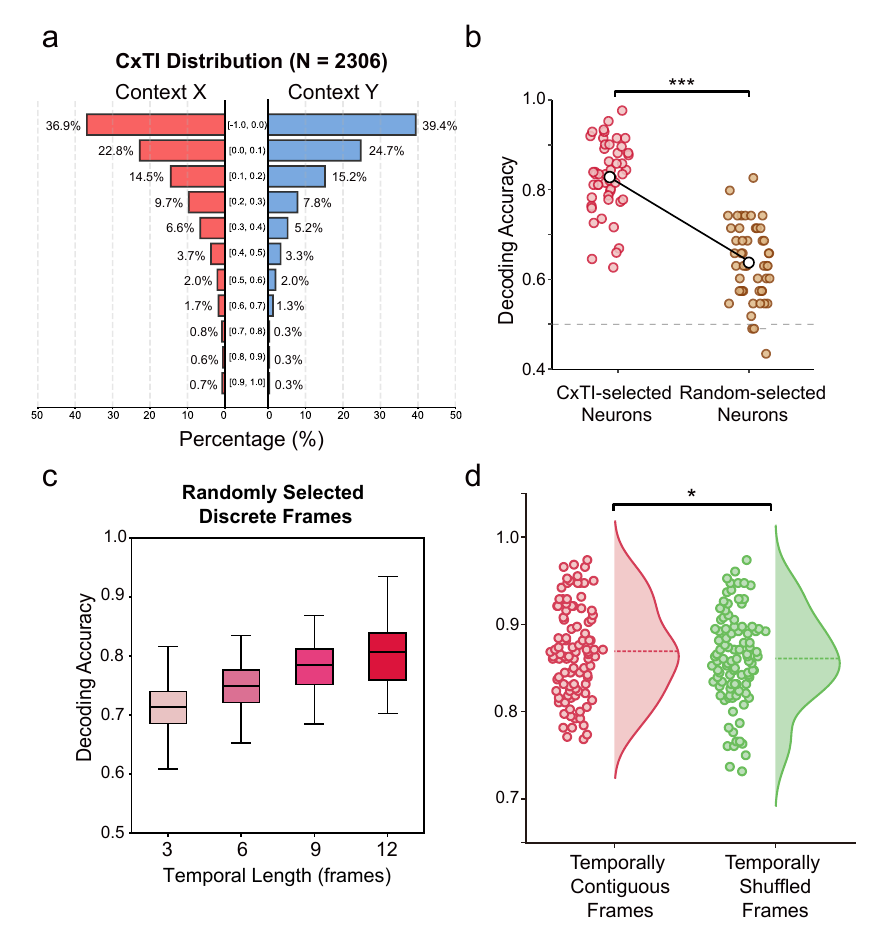}
\subsection*{Extended Data Fig.1: Complementary analyses of mouse mPFC activity further support context representations.}

(a) Distribution of the context-tuning index (CxTI) for n = 2,306 neurons recorded under context X (red; spatial-frequency rule) and context Y (blue; orientation rule). Bars indicate the percentage of neurons falling into each CxTI bin.
\\
(b) Linear SVM decoding accuracy using CxTI-selected neurons versus randomly selected neurons (matched neuron number). Each dot represents one random draw with a different random seed; white markers denote the mean across draws. The dashed line denotes chance-level performance. ***p $<$ 0.001.
\\
(c) Decoding accuracy of contexts X and Y using neural response sequences with different temporal lengths (2, 5, 10, and 14 time bins) during stimulus presentation, where temporal sequences are independently randomly sampled for each temporal length.
\\
(d) Comparison of decoding accuracy between contiguously sampled temporal sequences and temporally shuffled temporal sequences. Each dot represents one decoding run using a randomly subsampled neuronal population and a random initialization; for visualization clarity, only a random subset of runs is shown. Kernel density estimates illustrate the distribution of decoding accuracies computed from all runs. Decoding was performed using temporal windows containing the same sampled time points in both conditions, with one condition preserving the original temporal order and the other randomly shuffling their order. Contiguously ordered temporal sequences yielded significantly higher decoding accuracy than temporally shuffled sequences (*p $=$ 0.0114).
\newpage

\noindent\includegraphics[width=0.95\linewidth]{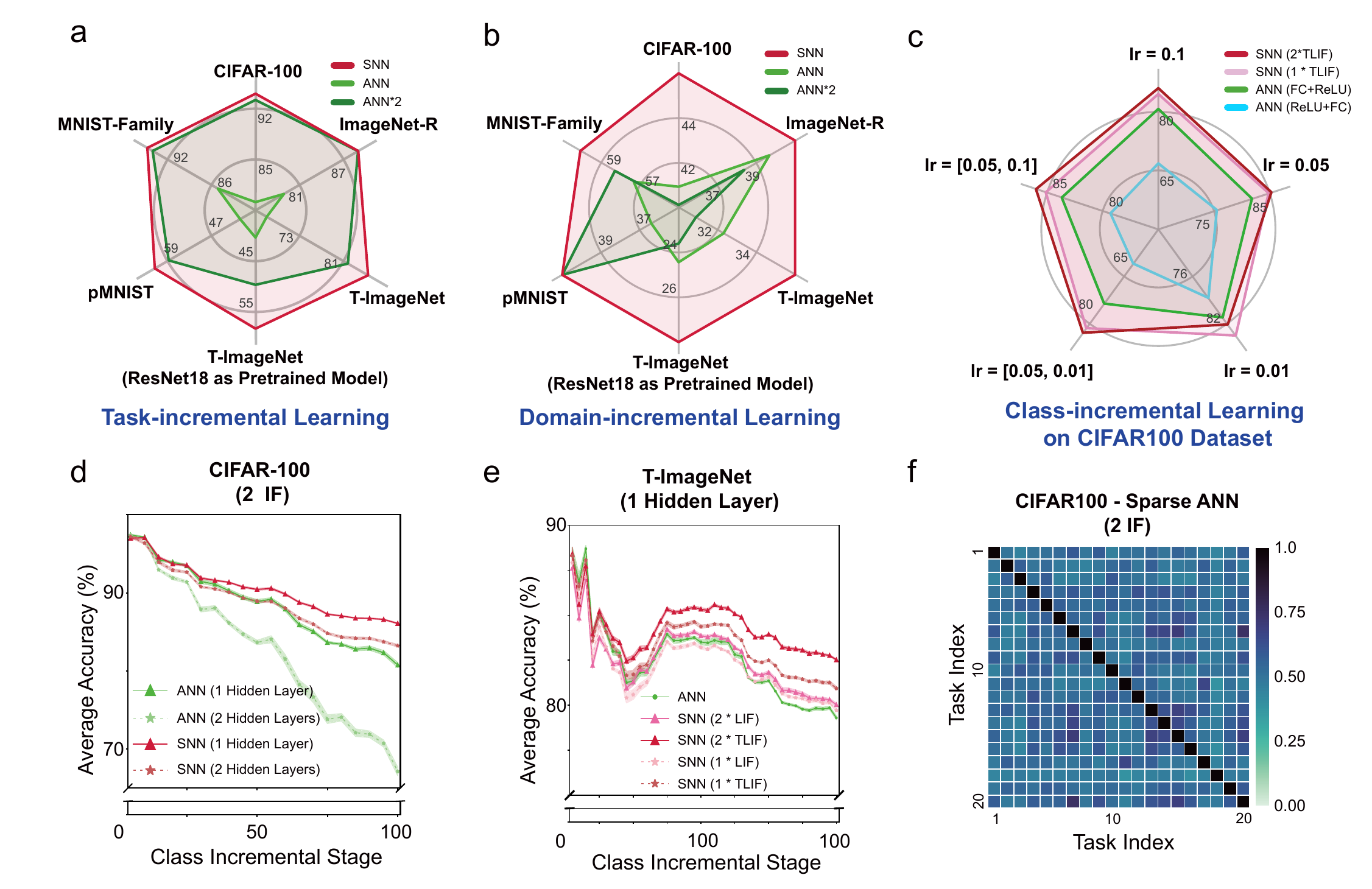}
\subsection*{Extended Data Fig.2: Comparison between different SNN configurations and ANNs in lifelong learning scenarios.}

(a, b) Radar plot results for (a) task-incremental learning and (b) domain-incremental learning across multiple datasets, including CIFAR-100, MNIST-family datasets, pMNIST, T-ImageNet, T-ImageNet with a ResNet pretrained model, and ImageNet-R.\\
(c) Results of the radar plot for class-incremental learning in the CIFAR-100 dataset across different setups of the learning rate (lr).\\
(d) Average accuracy across learned classes during class-incremental learning. Triangles indicate results from models with one hidden layer, and stars indicate results from models with two hidden layers.\\
(e) Average accuracy across learned classes during class-incremental learning. Triangles indicate models with multiple IF layers (2 $\times$ IF), whereas stars indicate models with a single IF layer after the input layer (1 $\times$ IF).\\
(f) Task-by-task overlap of high-activity subnetworks. Heat maps show the overlap fraction of the top 100 neurons ranked by firing rate between all pairs of tasks for sparse ANN (2 IF).
\newpage

\noindent\includegraphics[width=0.95\linewidth]{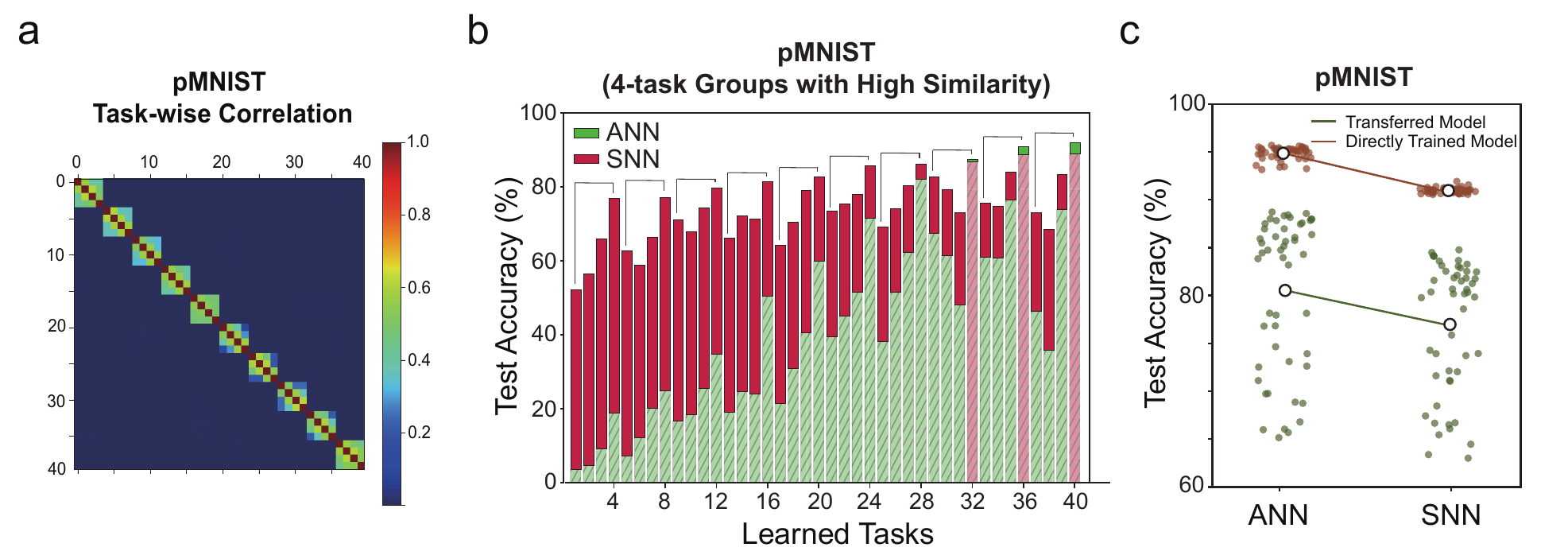}
\subsection*{Extended Data Fig.3: Transfer efficiency experiments on permuted MNIST dataset.}

(a) Correlation matrix across samples from 40 tasks in the pMNIST dataset. Tasks are organized into groups of four, with tasks within each group sharing similar features.\\
(b) Permuted MNIST (pMNIST) with a similarity structure: tasks are organized into 4-task groups with high within-group similarity. Bar plots show final per-task accuracy after learning all tasks for ANN (top) and SNN (bottom).\\
(c) Test accuracy on the last task within each group. The transferred model is trained on the first three tasks and evaluated on the last task, whereas the directly trained model is trained and evaluated on the last task directly. White dots denote the mean across runs.\\

\newpage

\noindent\includegraphics[width=0.95\linewidth]{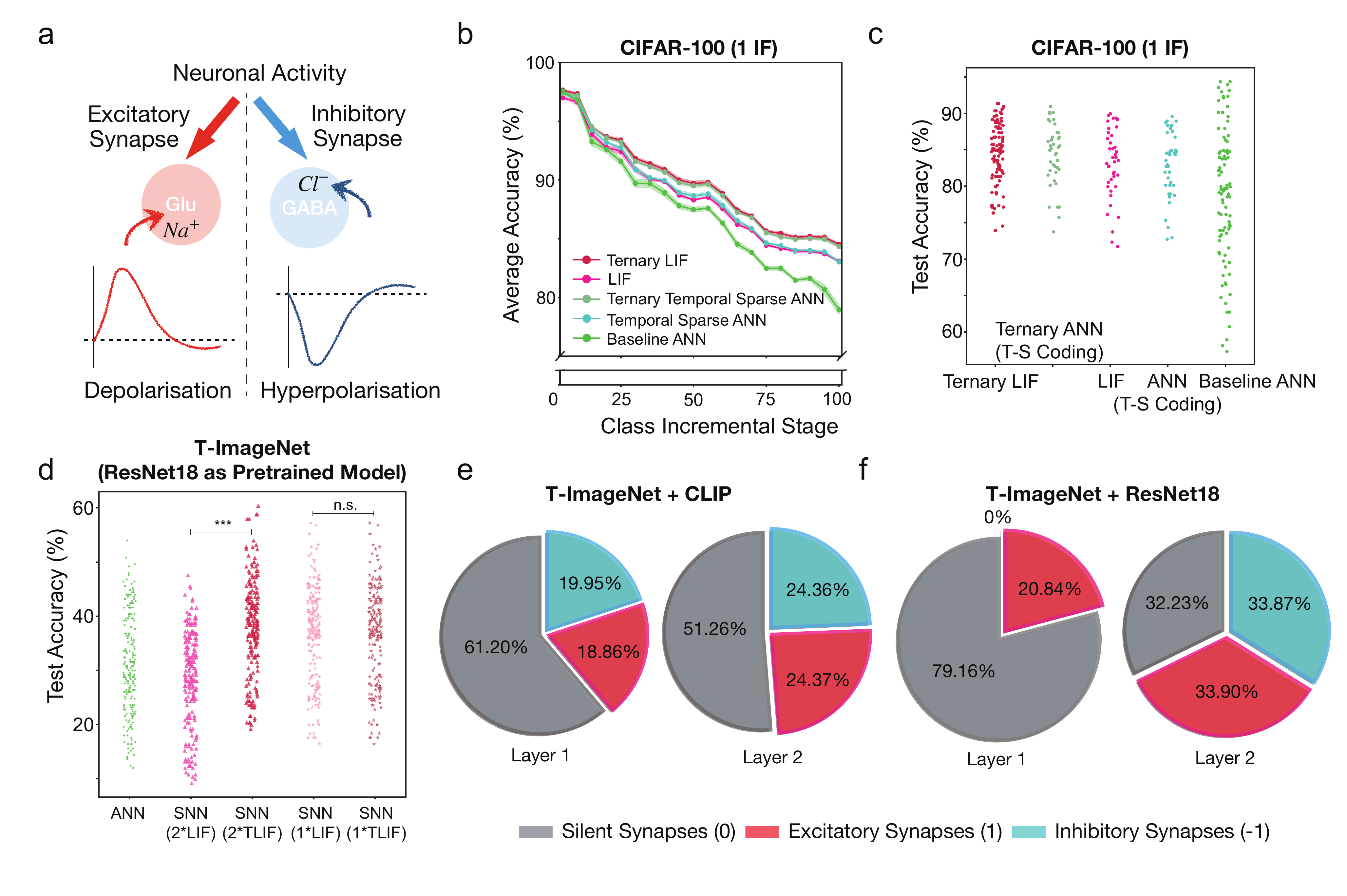}
\subsection*{Extended Data Fig.4: Inclusion of inhibitory states in spiking dynamics improves lifelong-learning performance.}

(a) Schematic illustration of postsynaptic responses evoked by excitatory and inhibitory synapses in the brain. Excitatory synaptic input (left) induces a positive-going, depolarizing membrane-potential response through glutamatergic signaling and Na$^{+}$ influx, whereas inhibitory synaptic input (right) induces a negative-going, hyperpolarizing membrane-potential response through GABAergic signaling and Cl$^{-}$ influx.\\
(b) Average test accuracy across class-incremental learning stages on CIFAR-100. Accuracy is averaged across learned classes at each incremental stage.\\
(c, d) Test accuracy distributions at the final learning stage on the (c) CIFAR-100 dataset and the (d) T-ImageNet dataset. Each dot represents one run; statistical significance is indicated (***p $<$ 0.001; n.s., not significant).\\
(e, f) Composition of neuronal populations in Layer~1 and Layer~2 of the IF neuron model for (e) T-ImageNet + CLIP and (f) T-ImageNet + ResNet18, showing the proportions of silent synapses (0), excitatory synapses (1), and inhibitory synapses ($-1$).\\

\newpage

\begin{center}
    \noindent\includegraphics[width=0.8\linewidth]{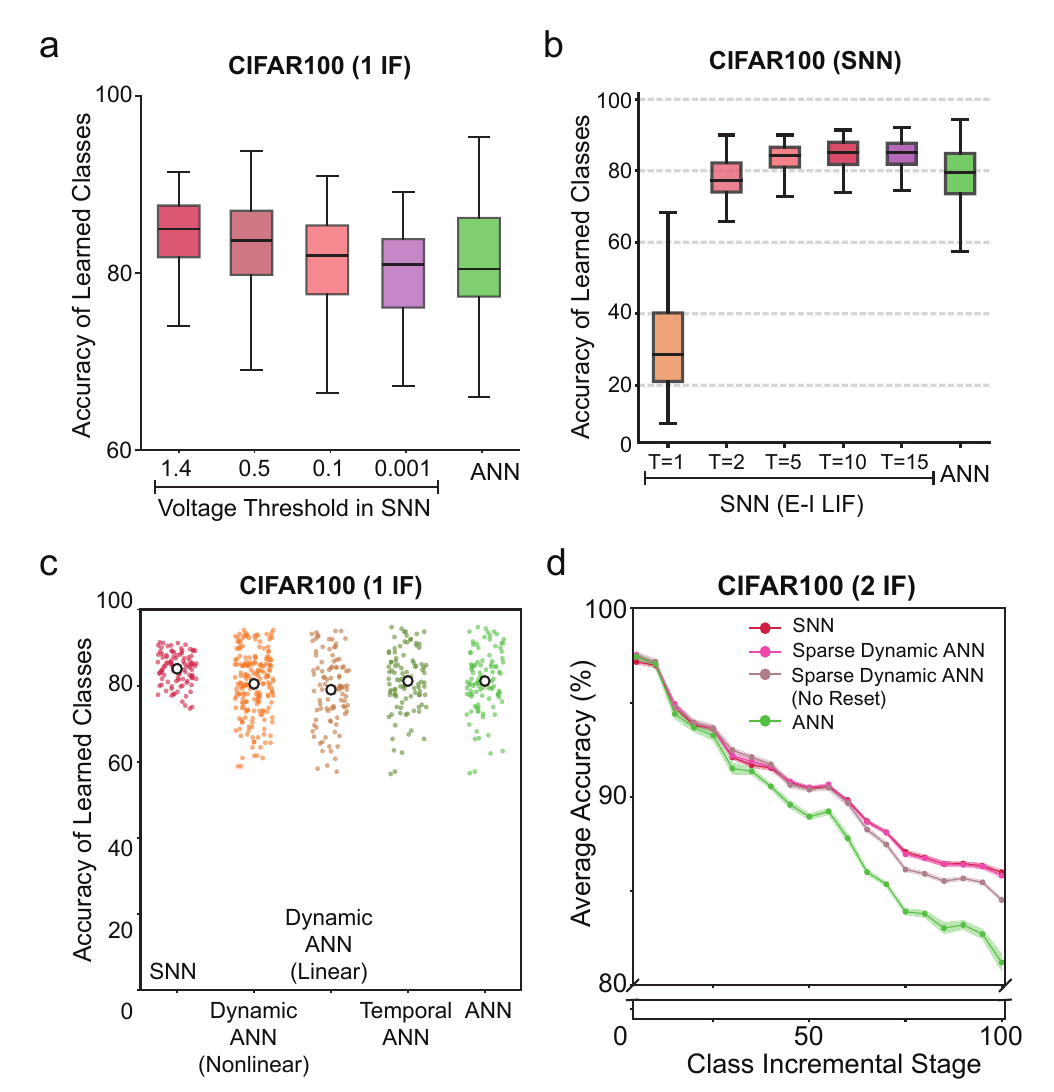}
\end{center}
\subsection*{Extended Data Fig.5: Evaluation of sparse and temporal coding mechanisms on computational models.}

(a) Accuracy of learned classes on CIFAR-100 for SNNs with different voltage thresholds (1 IF), compared with an ANN baseline.
\newline
(b) Box-and-whisker plots of last-stage average accuracy across random seeds across increasing time length (T) for SNN. Center lines indicate the median, boxes indicate the interquartile range (IQR), and whiskers indicate the range (minimum to maximum).
\newline
(c) Distribution of final-stage test accuracy across random seeds for various temporal coding ANN. 
\newline
(d) Average test accuracy across class-incremental stages on CIFAR-100 (2 IF, mean $\pm$ SEM across five random seeds).
\newpage

\begin{center}
    \includegraphics[width=0.85\linewidth]{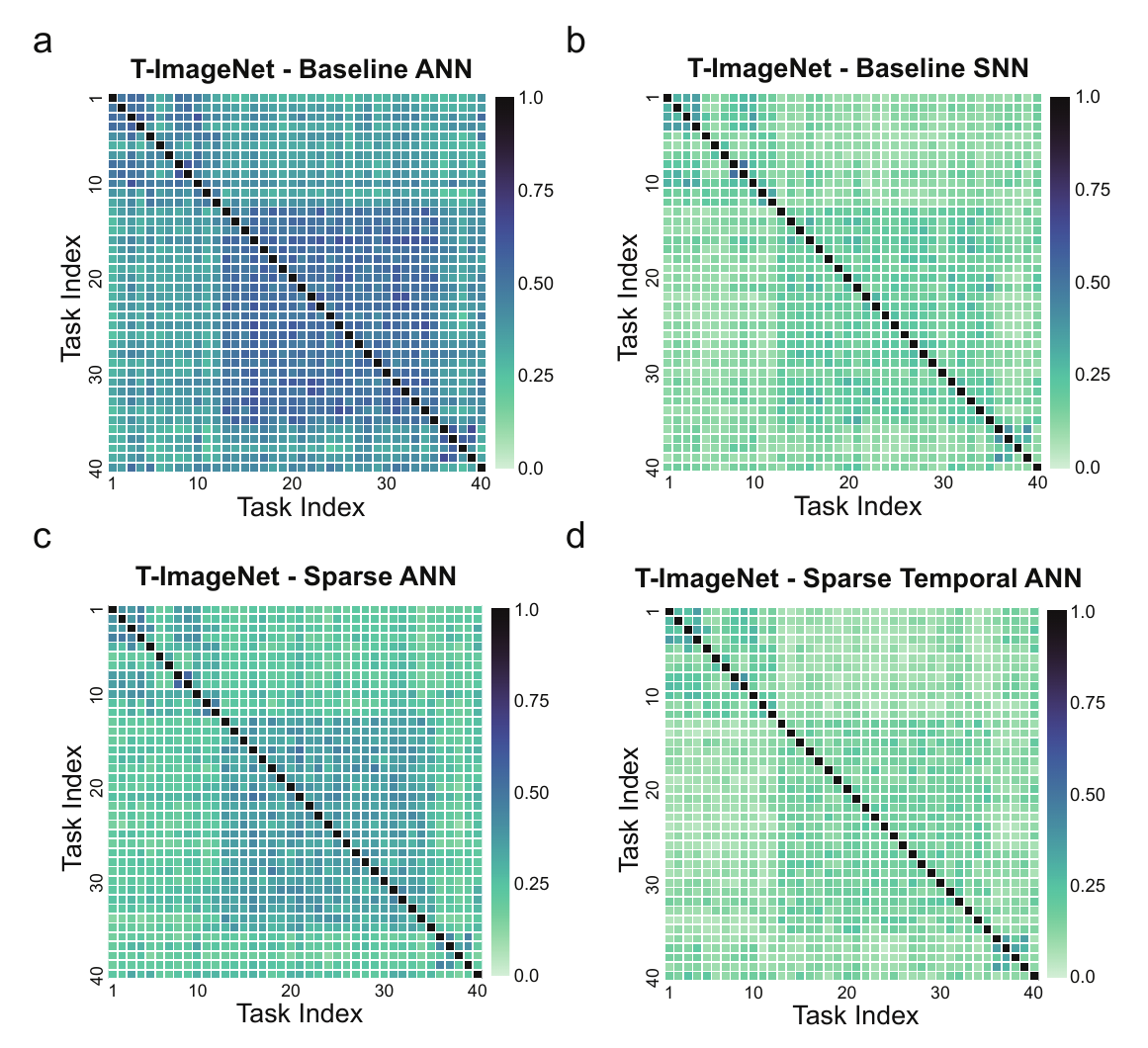}
\end{center}
\subsection*{Extended Data Fig.6: Task-by-task overlap of high-activity subnetworks on T-ImageNet dataset.}

Heat maps show the overlap fraction of the top 100 neurons ranked by firing rate between all pairs of tasks for (a) Baseline ANN, (b) Baseline SNN, (c) Sparse ANN, and (d) sparse temporal ANN.
\newpage

\end{appendices}
\end{document}